\documentclass[twocolumn,nofootinbib,amsmath,amssymb,aps,prd,balancelastpage,superscriptaddress]{revtex4-1}

\usepackage{amsmath,amssymb}
\usepackage{amsfonts,amssymb,mathrsfs}
\usepackage{graphicx}
\usepackage[bookmarks=false,pdfstartview=FitH]{hyperref}
\usepackage[all]{hypcap}
\usepackage{epsfig,latexsym}

\RequirePackage{xspace} \allowdisplaybreaks

\def\bef{\begin{figure}}
\def\eef{\end{figure}}

\newcommand{\be}[1]{\begin{equation}\label{#1}}
\newcommand{\beq}{\begin{equation}}
\newcommand{\ee}{\end{equation}}
\newcommand{\beqn}[1]{\begin{eqnarray}\label{#1}}
\newcommand{\eeqn}{\end{eqnarray}}
\newcommand{\bd}{\begin{displaymath}}
\newcommand{\ed}{\end{displaymath}}

\def\lsim{\raise0.3ex\hbox{$\;<$\kern-0.75em\raise-1.1ex
e\hbox{$\sim\;$}}}
\def\gsim{\raise0.3ex\hbox{$\;>$\kern-0.75em\raise-1.1ex
\hbox{$\sim\;$}}}
\def\simlt{\mathrel{\lower2.5pt\vbox{\lineskip=0pt\baselineskip=0pt
           \hbox{$<$}\hbox{$\sim$}}}}
\def\simgt{\mathrel{\lower2.5pt\vbox{\lineskip=0pt\baselineskip=0pt
           \hbox{$>$}\hbox{$\sim$}}}}
\def\unity{{\hbox{1\kern-.8mm l}}}

\renewcommand{\to}{\rightarrow}
\renewcommand{\vec}[1]{\mathbf{#1}}

\renewcommand{\to}{\rightarrow}

\def\lsim{\mathrel{\mathop  {\hbox{\lower0.5ex\hbox{$\sim$}
\kern-0.8em\lower-0.7ex\hbox{$<$}}}}}
\def\gsim{\mathrel{\mathop  {\hbox{\lower0.5ex\hbox{$\sim$}
\kern-0.8em\lower-0.7ex\hbox{$>$}}}}}

\def\be{\begin{equation}}
\def\ee{\end{equation}}

\usepackage{color}

\begin{document}

\pagestyle{plain}

\title{Mirror Symmetry of quantum Yang-Mills vacua and cosmological implications}

\author{Andrea Addazi}
\email{andrea.addazi@lngs.infn.it}
\affiliation{Department of Physics \& Center for Field Theory and Particle Physics, Fudan University, 
200433 Shanghai, China}

\author{Antonino Marcian\`o}
\email{marciano@fudan.edu.cn}
\affiliation{Department of Physics \& Center for Field Theory and Particle Physics, 
Fudan University, 200433 Shanghai, China}

\author{Roman Pasechnik}
\email{Roman.Pasechnik@thep.lu.se}
\affiliation{Department of Astronomy and Theoretical Physics,
Lund University, SE-223 62 Lund, Sweden}

\author{George Prokhorov}
\email{prokhorov@theor.jinr.ru}
\affiliation{Bogoliubov Laboratory of Theoretical Physics, 
Joint Institute for Nuclear Research, Dubna, Russia}

\begin{abstract}
\noindent
We find an argument related to the existence of a $\mathbb{Z}_2$-symmetry for the renormalization group flow derived from the bare Yang-Mills Lagrangian, and show that 
the cancellation of the vacuum energy may arise motivated both from the renormalization group flow solutions and the effective Yang-Mills action. In the framework of the effective Savvidy's action, 
two Mirror minima are allowed, with exactly equal and hold opposite sign energy densities. At the cosmological level, we explore the stability of the electric and magnetic attractor solutions, both within 
and beyond the perturbation theory, and find that thanks to these latter the cancellation between the electric and the magnetic vacua components is achieved at macroscopic space and time separations. 
This implies the disappearance of the conformal anomaly in the classical limit of an effective Yang-Mills theory. In this picture, the tunneling probability from the Mirror vacua to the other vacua is exponentially 
suppressed in the quantum non-thermal state --- similarly to what happens for electroweak instantonic tunneling solutions. Specifically, we show that, in a dynamical Friedmann-Lema\^itre-Robertson-Walker (FLRW) 
cosmological background, the Nielsen-Olsen argument --- on the instability of uniform chromo-electric and chromo-magnetic Mirror vacua --- is subtly violated. The chromo-magnetic and chromo-electric uniform 
vacua are unstable only at asymptotic times, but at those times the attractor to a zero energy density is already reached. The two vacua can safely decay into one anisotropic vacuum that 
has zero energy-density inside the Fermi confinement volume scale. We also discover a new surprising pattern of solitonic and anti-solitonic space-like solutions, which are sourced by the Yang-Mills dynamics coupled 
to the Einstein's equations in FLRW. We dub such non-perturbative configurations, which are directly related to dynamical cancellation mechanism of the vacuum energy, as {\it chronons}, or $\chi$-solutions. 
\end{abstract}

\maketitle
\noindent
\medskip


\section{Introduction}
\noindent
The ground state of quantum Yang-Mills (YM) theories plays a crucial role in both particle physics and cosmology. In particular the gluon condensate, a strongly coupled system in quantum chromo dynamics (QCD), largely determines non-trivial properties of the topological QCD vacuum. In a non-perturbative mechanism, this is responsible {\it e.g.} for the color confinement effects and the hadron mass generation --- for a comprehensive review on the topological QCD vacuum, see {\it e.g.}~Ref.~\cite{Shifman:1978bx} and references therein. Currently, the cosmological constant (CC) scenario, with the vacuum equation of state $w\equiv p/\epsilon=-1$, is preferred to the Dark Energy (DE) paradigm to unveil the late-time acceleration epoch, as supported by a wealth of data from the Supernovae type IA \cite{Reid:2009xm} and Cosmic Microwave Background \cite{Ade:2013zuv} observations. Despite of many DE/CC models existing in the literature, there is not a compelling resolution of the CC problem, {\it i.e.}~why the CC term is so small compared to the other scales of Nature, and why it is positive. 

From the Quantum Field Theory (QFT) viewpoint, the ground state energy density of the Universe should account for a bulk of various contributions from existing quantum fields, at energy scales ranging from the Quantum Gravity (Planck) scale, $M_{\rm PL}\simeq 1.2\cdot 10^{19}$ GeV, down to the QCD confinement scale, $\Lambda_{\rm QCD}\simeq 0.1$ GeV. Even the relatively well-known vacuum subsystems of the Standard Model (SM), such as the Higgs and the quark-gluon condensates, exceed by far the observed CC. This is often considered as a severe problem \cite{Zeldovich:1967gd,Weinberg:1988cp} --- for a recent review on this topic, see e.g.~Ref.~\cite{Pasechnik} and references therein. For confined QCD, with SU(3) color gauge symmetry, there is a rather unique contribution to the ground state energy of the Universe that emerges from the non-perturbative quantum-topological fluctuations of the quark and gluon fields \cite{Boucaud:2002nc,Hutter:1995sc,instantons,Shifman:1978bx}, namely, $\epsilon^{\rm QCD}_{\rm top} \simeq -(5\pm 1)\times 10^{9}\;\text{MeV}^4$. Given the fact that the CC term observed in astrophysical measurements is very small (and positive), $\epsilon_{\rm CC}\simeq 3\times 10^{-35}\,{\rm MeV}^4 \,,$ to the first approximation one must exclude the negative-valued topological vacuum contribution with an accuracy of a few tens of decimal digits.

Recently, in Ref.~\cite{Pasechnik:2016twe} by some of the authors it was shown that a possible elimination of the QCD contribution to the cosmological constant could be achieved by means of the existence of an additional ``Mirror QCD'' sector whose (non-perturbative) vacuum energy-density contributes with an opposite sign to the conventional QCD trace anomaly. Disregarding the Anthropic Principle, the main issue of this approach is the need for a significant fine tuning between the usual QCD and Mirror QCD vacua parameters which would be a problem for getting a naturally small CC term. Within this paper we show that even in the framework of standard QFT it is possible to recover as a result the cancellation of SU(2) Yang-Mills (YM) contributions to the vacuum energy within the {\it same} theory. This achievement holds a certain generality, since SU(2) subgroups of SU(N) YM theories can always be picked out, being the ones that must be accounted for the cosmological applications. The vacua compensation mechanism will be analyzed for effective YM theories, in both the perturbative and the non-perturbative cases, and then applied to address the QCD electric and magnetic condensates. Our approach is based on the Savvidy vacuum model \cite{Savvidy:1977as,Savvidy,Pagels,Dunne:2004nc}, as an effective method describing the ground state dynamics in quantum YM field theories at long distances. Interestingly enough, the Savvidy vacuum model has received a further support from another approach based on the analysis of the gluon condensation within the framework of the Functional RG (FRG) \cite{Eichhorn:2010zc,Dona:2015xia, Addazi:2016sot}. 

As the main result of this work, we find the stability conditions of the considered Savvidy vacuum solutions for the gauge-invariant homogeneous gluon condensate, and obtain analytic expressions for the density, the pressure and the scale factor in the non-stationary Friedmann-Lema\^itre-Robertson-Walker (FLRW) Universe filled with the gluon condensate, which fluctuates near the minimum of the effective Lagrangian.

\section{Effective YM Theory and the Mirror symmetry}
\noindent
We may start showing how to recover the effective action of SU(N) YM theories, following the seminal 
Refs.~\cite{Savvidy:1977as,Savvidy:1977as} recently followed by Refs.~\cite{Pasechnik:2013sga,Pasechnik:2013poa,Pasechnik:2013sga,Pasechnik:2013poa,
Prokhorov:2013xba,Pasechnik:2016sbh,Pasechnik:2016twe}. We then generalize these findings for a non-stationary FLRW background of expanding Universe. 

In order to incorporate the conformal anomaly via the variational procedure, the gauge coupling $g_{\rm YM}$ should acquire a dependence on the quantum fields, according to the RG equations. The order parameter of the theory is denoted with $\mathcal{J}$, a gauge-invariant operator of the least dimension \cite{Pagels}. In what follows, the running coupling constant $g_{\rm YM}$ recasts conventionally as $\bar{g}$, so to encode the dependence on $\mathcal{J}$ in the effective Lagrangian 
$\mathcal{L}_{\rm eff}$, namely,
\begin{eqnarray}
&& \mathcal{L}_{\rm eff}=\frac{\mathcal{J}}{4\bar{g}^2}\,, \quad  \bar{g}^2 = \bar{g}^2(\mathcal{J})\,, \quad 
\mathcal{J}=-\frac{\mathcal{F}_{\mu\nu}^{a}\mathcal{F}^{\mu\nu}_{a}}{\sqrt{-g}}\,, \label{L} 
\label{metric}
\end{eqnarray}
where $g\equiv {\rm det}(g_{\mu\nu})$, $g_{\mu\nu}=a(\eta)^2\mathrm{diag}(1,\,-1,\,-1,\,-1)$ is the FLRW metric, $\mathcal{A}_{\mu}^{a}$ are the SU(N) connections and $\mathcal{F}^a_{\mu\nu}$ -- their field-strength. Through the paper $a,\,b,...$ denote internal indices of SU(N) in the adjoint representation. 

For FLRW metrics $\mathcal{J}$ simplifies into 
$$\mathcal{J}=\frac{2}{\sqrt{-g}}\,\sum_a ( \vec{E}_a\cdot \vec{E}_a-\vec{B}_a\cdot \vec{B}_a)\equiv \frac{2}{\sqrt{-g}}\,(\vec{E}^2-\vec{B}^2)\,,$$ 
which is cast in terms of the electric field $\vec{E}_a$ and the magnetic field $\vec{B}_a$ components. We define the spatial average quantity $\langle \mathcal{J} \rangle$, and distinguish the cases in which: i) $\langle \mathcal{J} \rangle$ is positive, meaning that the average chromo-electric (CE) components $\langle \vec{E}^{2} \rangle$ dominate over the averaged chromo-magnetic (CM) terms $\langle \vec{B}^{2} \rangle$;  ii) {\it viceversa}, the case of a chromo-magnetically dominated state $\langle J \rangle <0$ corresponds to a CM condensate.

Through the rest of the paper we will work only with spatially averaged quantities, thus from now on we remove the $\langle \dots\rangle$, for simplicity. Our approach must be thought as a chromo-dynamical mean field theory, in analogy to many condensed matter models\footnote{For example, the Ginzburg-Landau model describes the evolution of spatially averaged observables in superconductive materials, which in turn are crystals with local impurities and anisotropies --- see {\it e.g.}~Ref.~\cite{GL}. }. In the minimum of the effective Lagrangian, the spatially-homogeneous CE and CM condensates correspond to positive- 
and negative-valued energy densities, respectively. In a non-stationary background of expanding Universe, these condensates yield stable de-Sitter (dS) and 
anti-de-Sitter (AdS) attractor solutions with positive and negative cosmological constants, respectively.

The gauge coupling satisfies the RG equation 
$$2\mathcal{J}\, \frac{d\bar{g}^{2}}{d\mathcal{J}}=\bar{g}^{2}\beta\,,$$
where $\beta=\beta(\bar{g}^{2})$ and the running of the coupling constant $\bar{g}^2$ is determined by the {\it exact} $\beta$-function ---  both the quantities can be either positive or negative, in general.

By the standard variational procedure, starting from the effective action (\ref{L}) we arrive at the all-loop effective YM equations of motion, supplemented by the RG equation, which can be represented as follows
\begin{align}
\label{EOM}
&\overrightarrow{\mathcal{D}}^{ab}_{\nu}\left[\frac{\mathcal{F}_{b}^{\mu\nu}}{\bar{g}^{2}\sqrt{-g}}
\left(1-\frac{\beta(\bar{g}^2)}{2}\right)\right]=0\,, \\
&\overrightarrow{\mathcal{D}}^{ab}_{\nu}\equiv \Big(\delta^{ab}\frac{\overrightarrow{\partial}_{\nu}\sqrt{-g}}{\sqrt{-g}} -  f^{abc}\mathcal{A}_{\nu}^{c}\Big)\, , \\
& \frac{d\ln |\bar{g}^2|}{d\ln |\mathcal{J}|/\mu_0^{4}}=\frac{\beta(\bar{g}^2)}{2} \,,\label{RGE-mod}
\end{align}
where $\mu_0$ is a scale parameter. Thus, for the system of 
equations (\ref{EOM}), we find the exact (partial) ground-state solution
\begin{eqnarray}
\label{ground-state}
\beta(\bar{g}_*^2)=2\,, \qquad \bar{g}_*^2\equiv \bar{g}^{2}(\mathcal{J}^*) \,, \qquad \mathcal{J}^*>0 \,,
\end{eqnarray}
which we refer to the {\it CE condensate}, in what follows. Is this the only possible ground state solution in a YM theory?

\section{Mirror symmetry }
\noindent
The effective YM Lagrangian (\ref{L}) in a vicinity of the ground state $\mathcal{J}\simeq \mathcal{J}^*$ is
$\mathbb{Z}_2$-symmetric w.r.t. {\it simultaneous} permutations
\begin{eqnarray} \nonumber
\mathbb{Z}_2: &\quad&  \mathcal{J}^* \, \longleftrightarrow  \, -\mathcal{J}^* \,, \quad
\bar{g}_*^2 \,  \longleftrightarrow  \, -\bar{g}_*^2\,, \\
&&\quad\quad \beta(\bar{g}_{*}^{2}) \, \, \longleftrightarrow  \, -\beta(\bar{g}_{*}^{2})\,,
\label{Z2-RG}
\end{eqnarray}
This important symmetry property hold {\it only} in the ground state of the effective YM theory and 
has relevant consequences on the stability of the ground-state YM solutions in the FLRW spacetime. 
Note, the $\mathbb{Z}_2$ symmetry (or Mirror symmetry, in what follows) effectively ``maps'' the CE condensate solution with 
$\mathcal{J}^*>0$ found in Eq.~(\ref{ground-state}) to another, CM condensate solution 
$\mathcal{J}^*<0$, and {\it vice versa}. In fact, due to the fact that the effective 
Lagrangian Eq.~(\ref{L}) is invariant under the $\mathbb{Z}_2$ symmetry in a vicinity of its minimum, 
the CE ($\mathcal{J}^*>0$) and the CM ($\mathcal{J}^*<0$) vacua should be associated with 
two equal (Mirror) minima of the effective Lagrangian.

We emphasize that this symmetry, which reveals itself {\it only} in the ground state, does not 
explicitly show itself in the equation of motion (\ref{EOM}). The CE condensate corresponds to a ground-state solution of 
the Eq.~(\ref{EOM}), which is satisfied for $\beta(\bar{g}_*^2)=2$. On the other hand, by means of the Mirror 
transformation (\ref{Z2-RG}),the CM vacuum 
corresponds to $\beta(\bar{g}_*^2)=-2$, which does not imply the vanishing of $[1-\beta(\bar{g}_*^2)/2]$, 
as for the CE case, but rather amounts to a $2$ overall factor in Eq.~(\ref{EOM}) which drops out. 
The CM vacuum is then obtained as a more complicated solution of  Eq.~(\ref{EOM}), which recasts the following equation 
\begin{align}
\label{EOM2}
\overrightarrow{\mathcal{D}}^{ab}_{\nu}\left[\frac{\mathcal{F}_{b}^{\mu\nu}}{\bar{g}^{2}\sqrt{-g}}\right]=0\,,
\end{align}
resembling the standard classical YM equation of motion in a vicinity of the ground state typically studied 
in e.g. the instanton theory. 

Also, considering the energy-momentum tensor (EMT) associated to the two minima, the symmetry does not appear explicitly. 
The EMT of the Savvidy's theory reads
\begin{equation}
\label{energyMomentum2}
T_{\mu}^{\nu}=\frac{1}{\bar{g}^{2}}\Big[\frac{\beta(\bar{g}^{2})}{2}-1\Big]\Big(\frac{\mathcal{F}_{\mu\lambda}^{a} 
\mathcal{F}^{\nu\lambda}_{a}}{\sqrt{-g}}+\frac{1}{4}\delta_{\mu}^{\nu}\mathcal{J}\Big)-\delta_{\mu}^{\nu}\frac{\beta(\bar{g}^{2})}{8\bar{g}^{2}}\mathcal{J}\, . 
\end{equation}
In the case of the CE vacuum, the EMT simplifies to the trace-form 
\begin{equation}
\label{energyMomentum3}
T_{\mu}^{\mu}=-\frac{\beta(\bar{g}_{*}^{2})}{2\bar{g}_{*}^{2}}\mathcal{J}^{*}=-\frac{1}{\bar{g}_{*}^{2}}\mathcal{J}^{*}\, . 
\end{equation}
For the CM vacuum case, the EMT appears more complicated: 
\begin{equation}
\label{energyMomentum4}
T_{\mu}^{\nu}=\frac{-2}{\bar{g}^{2}}\Big(\frac{\mathcal{F}_{\mu\lambda}^{a} \mathcal{F}^{\nu\lambda}_{a}}{\sqrt{-g}}+
\frac{1}{4}\delta_{\mu}^{\nu}\mathcal{J}^{*}\Big)-\delta_{\mu}^{\nu}\frac{\beta(\bar{g}_{*}^{2})}{8\bar{g}_{*}^{2}}\mathcal{J}^{*}\, . 
\end{equation}
However, if we consider its trace, we obtain exactly the same trace-tensor of the CE vacua, but with an opposite sign: 
\begin{equation}
\label{energyMomentum5}
T_{\mu}^{\mu}=\frac{1}{\bar{g}_{*}^{2}}\mathcal{J}^{*}\, .  
\end{equation}
In fact, this could be noticed already at the level of generic EMT in Eq.~(\ref{energyMomentum2}) where 
the trace of the expression in the brackets is equal identically to zero such that only the last term contributes to the EMT trace.

Remarkably, the two Mirror minima of the effective Lagrangian 
have an opposite energy density, which is found to be
\begin{eqnarray} \label{eps-vac}
\epsilon_{\rm vac} \equiv \frac{1}{4}\langle T_\mu^\mu\rangle_{\rm vac} =  
\mp \mathcal{L}_{\rm eff}(\mathcal{J}^*) \,,
\end{eqnarray}
where the upper sign corresponds to the CE condensate, the lower --- to the CM condensate. Indeed, the 
$\mathcal{J}^{*}\leftrightarrow -\mathcal{J}^{*}$ transformation corresponds to an exchange of the electric and the magnetic components, 
which flip simultaneously the sign of the $\beta$-function and $\bar{g}_*^2$ for a fixed minimal (negative) value of $\mathcal{L}_{\rm eff}$.

The Perturbation Theory can be applied to the effective action in the limit of large mean fields, i.e. $|\mathcal{J}|\to \infty$, away from 
the nonperturbative ground state. We now comment on the one-loop results obtained by Savvidy for SU(N) YM theories, and then focus on 
a different strategy to account for all-loops corrections, based on the FRG approach. The latter has been developed 
in its cosmological applications in Ref.~\cite{Dona:2015xia}, accounting for the SU(2) gauge symmetry.

The standard one-loop SU(N) $\beta$-function reads
\begin{eqnarray}
\label{beta-g2-1L}
\beta_{1}=-\frac{bN}{48\pi^2}\,\bar{g}_{1}^2 \,, \qquad b=11 \,,
\end{eqnarray}
and the corresponding solution of the RG equation (\ref{RGE-mod}) is given by
\begin{eqnarray}
\bar{g}_{1}^2(\mathcal{J}) = \frac{\bar{g}_{1}^2(\mu_0^4)}{1+\frac{bN}{96\pi^2}\bar{g}_{1}^2(\mu_0^4)
\ln(|\mathcal{J}|/\mu_0^{4})} \,.
\label{RG-sol-1L}
\end{eqnarray}
Taking the position of the minimum of the effective Lagrangian as the physical scale of 
the considering quantum YM theory, i.e.
\begin{eqnarray}
\mu_0^4 \equiv |\mathcal{J}^*| \,,
\end{eqnarray}
we observe that indeed $\mathbb{Z}_2$ symmetry (\ref{Z2-RG}) is a symmetry of the ground state only. 

Note, for one of the two possible branches related by the Mirror symmetry (\ref{Z2-RG}), for which $\bar{g}_{1}^2(\mathcal{J}^*) > 0$,
the RG solution (\ref{RG-sol-1L}) can be conventionally rewritten as
\begin{eqnarray} \label{g1}
\bar{g}_{1}^2(\mathcal{J}) = \frac{96\pi^2}{bN\ln(|\mathcal{J} |/\lambda^4)} \,,
\end{eqnarray}
where
\begin{eqnarray}
\lambda^4 \equiv |\mathcal{J}^*|\exp\Big[-\frac{96\pi^2}{bN\bar{g}_{1}^2(\mathcal{J}^*)}\Big] \,,
\label{scale}
\end{eqnarray}
and the ground-state value of the coupling $\bar{g}_{1}^2(\mathcal{J}^*)$ can be either positive or negative depending on 
the considering branch. Thus, the corresponding one-loop effective action for the pure SU(N) gauge theory takes the following form
\begin{eqnarray}
\mathcal{L}^{(1)}_{\rm eff} = \frac{bN}{384\pi^2}\,\mathcal{J} \ln\Big(\frac{|\mathcal{J} |}{\lambda^4}\Big)\,,
\label{L-1L}
\end{eqnarray}
such that one recovers the well-known results obtained by Savvidy in Ref.~\cite{Savvidy:1977as}. 
Note, due to the Mirror $\mathbb{Z}_2$ symmetry, the CM and CE condensates correspond to the Mirror minima 
with the same value of the effective Lagrangian. Finally, the UV renormalisation is not affected by the Mirror symmetry 
since both RG solutions for $\bar{g}_{1}^2(\mathcal{J})$ related by the mirror symmetry have exactly the same form as 
in Eq.~(\ref{g1}) but having different physical scales $\lambda_{\pm}$
\begin{eqnarray}
\lambda_{\pm}^4 = |\mathcal{J}^*|\exp\Big[\mp\frac{96\pi^2}{bN|\bar{g}_{1}^2(\mathcal{J}^*)|}\Big] \,,
\label{scalep}
\end{eqnarray}
such that the two-scale structure of the gluonic vacuum is manifest.
\begin{figure}[!t]
 \centerline{\includegraphics[width=0.45\textwidth]{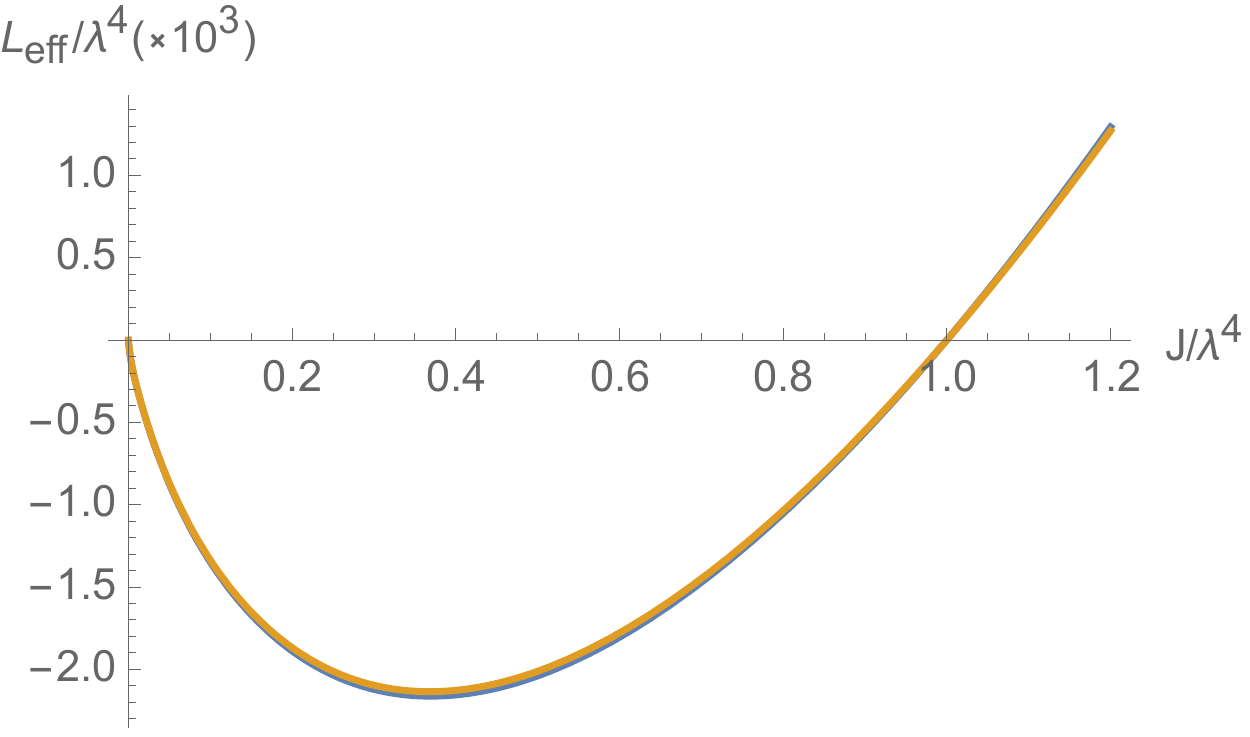}}
   \caption{ \small 
The effective SU(2) YM theory Lagrangian dependence on $\mathcal{J}/\lambda^4$ corresponding to one particular (chromoelectric) branch of 
the RG equation (\ref{RGE-mod}) with $\mathcal{J}>0$. The curves corresponding to the one-loop and all-loop effective Lagrangians
are practically indistinguishable.}
 \label{fig:Leff-CE}
\end{figure}

In Fig.~\ref{fig:Leff-CE}, we show the effective SU(2) YM theory Lagrangian dependence on $\mathcal{J}/\lambda^4$ corresponding 
to one particular branch of the RG equation (\ref{RGE-mod}) with $\mathcal{J}>0$ (CE configuration). As anticipated, there is a single minimum in the non-perturbative
domain $0<\mathcal{J}^*<\lambda^4$, hence, identified with the CE condensate. The Mirror CM condensate solution can then be obtained
by means of $\mathbb{Z}_2$ transformations (\ref{Z2-RG}), and it corresponds to the conventional one-loop 
result for the trace anomaly in SU(N) YM gluodynamics (known e.g. from lattice QCD simulations). Notably, applying the 
$\mathbb{Z}_2$ transformations to the physical scale of the CE configuration (\ref{scale}) one gets a smaller physical scale of the
CM configuration, such that the corresponding CM condensate $\mathcal{J}^*<0$ appears in the perturbative 
$|\mathcal{J}^*|>\lambda^4$ domain, with a positive $\bar{g}_{1}^2(\mathcal{J}^*) > 0$.

How well the one-loop approximation reproduces the all-loops vacuum state, given by the non-perturbative ground-state solutions in Eq.~(\ref{ground-state})? We can answer 
this question focusing on the case of SU(2), which is also relevant for cosmology, in the framework of FRG \cite{Eichhorn:2010zc,Dona:2015xia, Addazi:2016sot}. As is illustrated explicitly by two curves in Fig.~\ref{fig:Leff-CE} for the CE branch, the one-loop and the all-loops CE solutions 
approach the zero of the effective action at {\it exactly} the same values of $\mathcal{J}=0$ and $\mathcal{J}=\lambda^4$. The solutions also 
exhibit minima that, although do not coincide, are very close to each other: at one loop, ${|\mathcal{J}^*|}/{\lambda^4} = \frac{1}{e} \simeq 0.3679 \,, $ and ${\mathcal{L}^*_{\rm eff}}/{\lambda^4} = 
\! \pm {b}/(192\pi^2 e) \! \simeq \pm 2.135\cdot 10^{-3}$; at all loops $ {|\mathcal{J}^*|}/{\lambda^4} \simeq 0.3693 \,, $ and ${\mathcal{L}^*_{\rm eff}}/{\lambda^4} = \pm 2.163\cdot 10^{-3} \!.$ 
Remarkably, the CE ground-state solutions for one-loop and all-loops cases differ only at a per-mille level. By means of the Mirror symmetry, the same applies for the CM configuration as well.

It is worth emphasizing that is not reductive to focus on SU(2) YM theory. For any SU(N) gauge group, the cosmological instantiation will be provided by the SU(2) subgroups, for which an isomorphism 
between indices of the adjoint representation and spatial indices may be recovered. On the other hand, the calculation of the super-trace would be technically very difficult to be achieved. Because of the lack 
of any physical advantage, we can skip this point without any loss of generality and physical insight.

As the bottomline of this consideration, for the two Mirror vacua found from Eq.~\eqref{ground-state}, the net energy density gets both CM (perturbative) and CE (nonperturbative) vacua contributions 
with an equal modulus but an opposite sign which therefore cancel out
\begin{eqnarray}
&\epsilon^{\rm CE}_{\rm vac}\big|_{\mathcal{J}^*>0}+\epsilon^{\rm CM}_{\rm vac}\big|_{\mathcal{J}^*<0}\equiv 0\,,
\end{eqnarray}
{\it if and only if} both vacua do co-exist in the ground state of the Universe. We notice that this statement is valid both in one-loop and all-loops cases. From such a simple argument the vacuum energy-density 
cancellation may be envisaged. In the case of strongly-coupled SU(3) gluodynamics, such a cancellation is expected to happen beyond the confinement length-scale which would automatically yield 
vanishing mean-fields of gluons at large distances (when averaged over macroscopic volumes). The co-existence of the vacua in the quantum ground state thus implies their mutual screening, 
yielding a vanishing CC term in consistency with cosmological observations. 

\section{Homogeneous YM condensates}
\noindent
A gauge-invariant description of spatially homogeneous isotropic YM condensates, which depend only on time, can be obtained, assuming the gauge condition $A^a_0=0$. Due to the local isomorphism of the isotopic SU(2) gauge group and the SO(3) group of spatial 3-rotations, the unique (up to a rescaling) SU(2) YM configuration can be parameterized in terms of a scalar time-dependent spatially-homogeneous field --- see e.g. 
Refs.~\cite{KP1,KMPR,Cervero:1978db,Henneaux:1982vs, Hosotani:1984wj}. Within the symmetric gauge, one obtains a unique and gauge-invariant decomposition of 
the gauge field into a spatially homogeneous isotropic part (the YM condensate) and a non-isotropic/non-homogeneous parts (the YM waves), namely, 
$$A_{ak}\big(t,\vec{x}\big) = \delta_{ak} U(t) + \widetilde{A}_{ak}\big(t,\vec{x}\big)\,,$$ 
with 
$ \langle \widetilde{A}_{ik}\big(t,\vec x\big) \rangle = \int d^3x\; \widetilde{A}_{ik}\big(t,\vec x\big) = 0 $ 
and the YM condensate positively definite $U(t)>0$. In the QFT formulation, the inhomogeneous 
YM wave modes $\widetilde{A}_{ik}$ are interpreted as YM quanta (e.g. gluons), while $U(t)$ contributes to the ground state of the theory --- for further technical details, see Appendix~\ref{Sec:EYM}.

We may now focus on the equations of motion, addressing the time evolution of the homogeneous YM condensate in the cosmological environment. For this purpose, we consider the perturbative (one-loop) 
effective toy-model, provided that the exact (all-loop) formulation provides very similar results. In full analogy to the SU(2) condensate case \cite{Pasechnik:2013sga}, in the QCD case the system of the dynamical equations of the condensate has the exact solution corresponding to the vanishing logarithm or, equivalently, satisfies the transcendent equation $|Q|=1$, with 
\begin{eqnarray}
Q &\equiv& \frac{32}{11} \pi^2 e{(\xi \Lambda_{\rm QCD})^{-4}}T^{\mu}_{\mu}[U] \nonumber\\
&=& {6e\Big[(U')^2-\frac{1}{4}U^4\Big]}{a^{-4}(\xi \Lambda_{\rm QCD})^{-4}}\,, \nonumber
\end{eqnarray}
which yields the two distinct cases $Q=\pm 1$ --- for more details, see {\it e.g.} Appendix~\ref{Sec:EYM}.

As was mentioned above, quite naturally, the exact compensation of the positive- and negative-valued gluon condensate contributions to the QCD ground state energy density would be realized, in particular, if both the electric and magnetic components $Q=\pm 1$ co-exist in the ground state of the Universe. At macroscopic distances the two contributions cancel, without any fine-tuning of the model parameters, due to their (time) attractor nature at large physical times. Within this hypothesis, both QCD subsystems should be generated during the cosmological QCD phase transition, and asymptotically acquire the same absolute values of the energy density, with opposite signs that trigger cancellation at large $t$ for arbitrary values of the normalization parameter $\xi$. 

\onecolumngrid


\begin{figure}[!h]
\begin{minipage}{0.32\textwidth}
 \centerline{\includegraphics[width=1.0\textwidth]{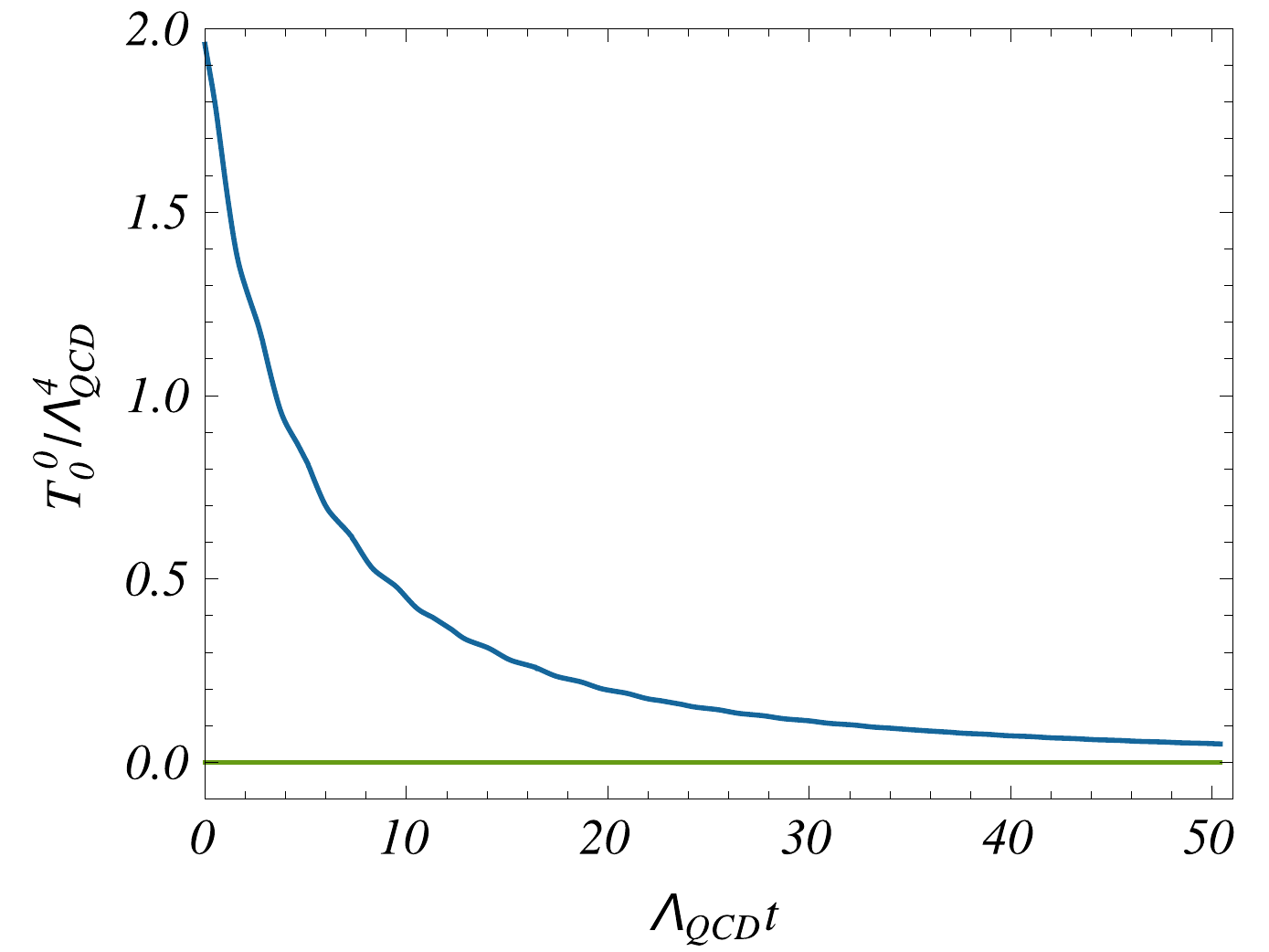}}
\end{minipage}
\begin{minipage}{0.32\textwidth}
 \centerline{\includegraphics[width=1.0\textwidth]{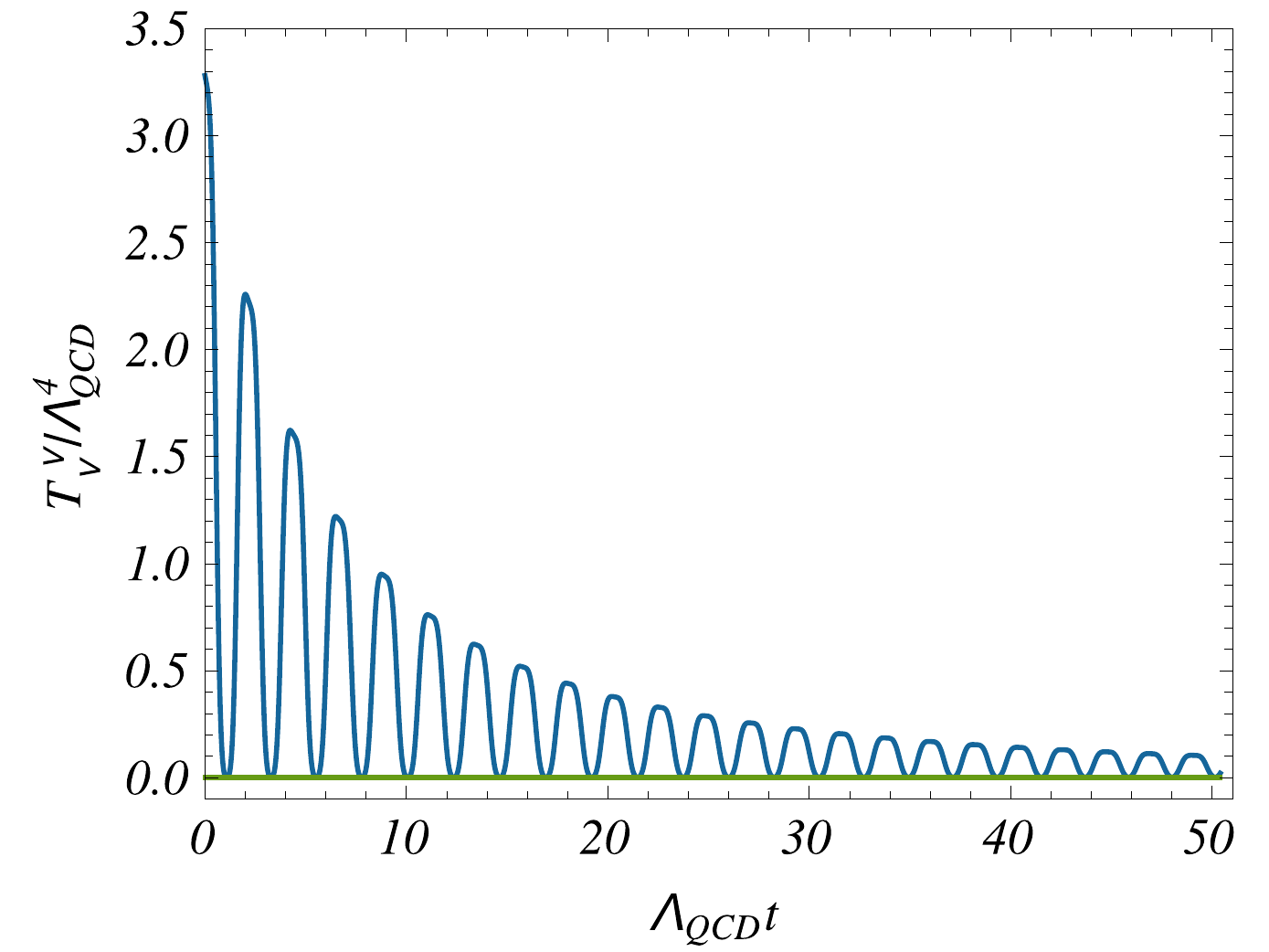}}
\end{minipage}
\begin{minipage}{0.32\textwidth}
\centerline{\includegraphics[width=1.0\textwidth]{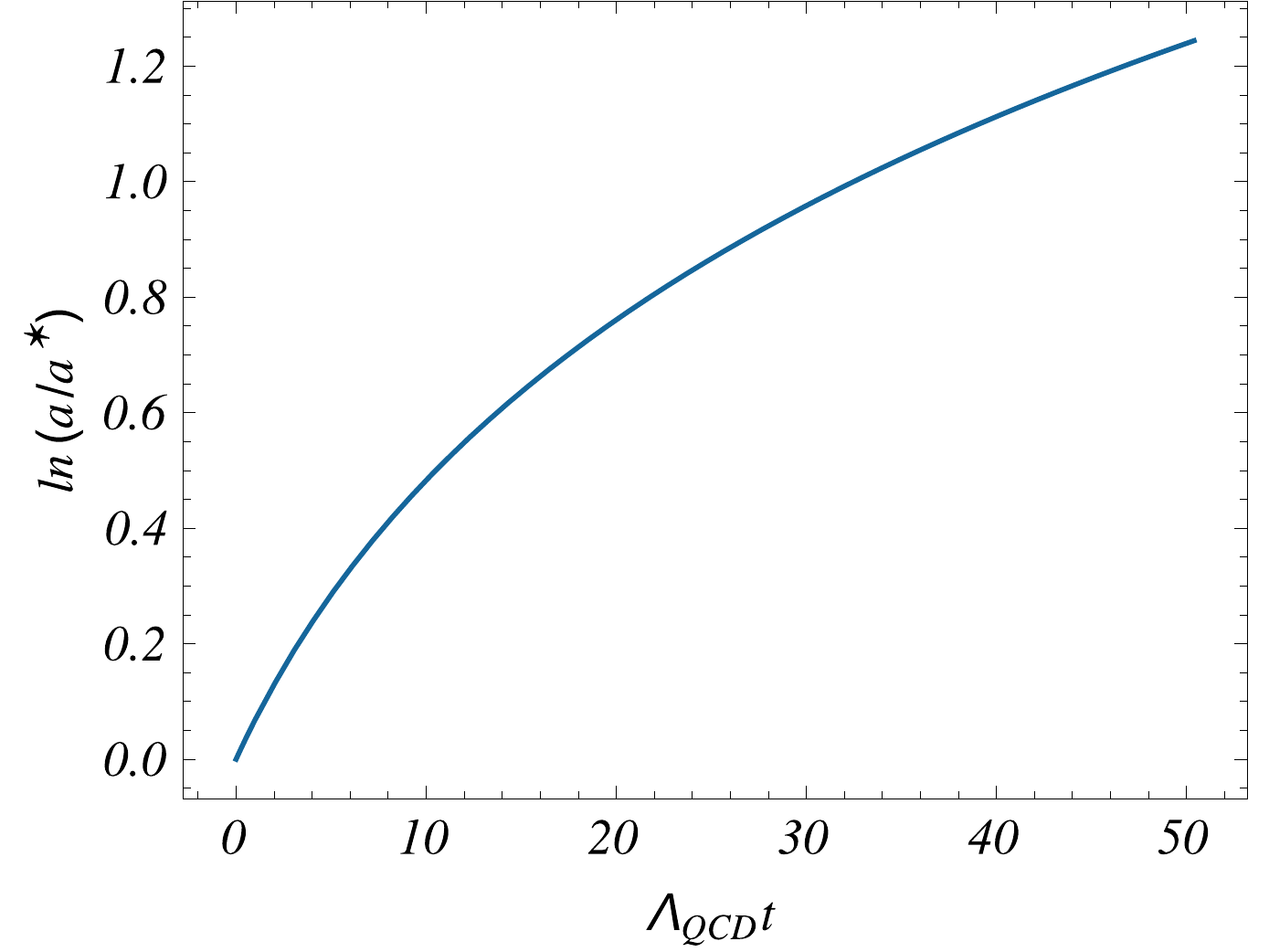}}
\end{minipage}
\caption{The total energy density $T^0_0(t)$ of the homogeneous gluon condensate (left), the trace of the total QCD energy-momentum tensor $T^\mu_\mu(t)$ (middle) 
and the logarithm of the scale factor $a(t)$ (right), are illustrated as functions of the physical time $t=\int ad\eta$ and in units of the characteristic time scale 
$\Lambda_{\rm QCD}^{-1}$. The total energy density and the trace values for $Q_0\equiv Q(t_0)=1$ are indicated by horizontal lines in the left and middle panels, 
respectively. Here, the initial conditions are chosen as $U_0=0$, $\dot{U}_0=(\xi \Lambda_{\rm QCD})^2/\sqrt{3e}$, $Q_0>1$, $\xi\simeq 4$, and 
the gravitational constant is set to $\varkappa=10^{-7} {\rm MeV}^{-2}$, for simplicity of the numerical analysis. Both quantities $T^0_0(t)$ and $T^\mu_\mu(t)$ 
are plotted in dimensionless units, and thus are rescaled by $\Lambda_{\rm QCD}^4$. The amplitude of the quasi-periodic oscillations of $Q=Q(t)$ decreases at large 
$t\gg \Lambda_{\rm QCD}^{-1}$, and asymptotically approaches unity, corresponding to the partial dS solution of the equations of motion.}
\label{fig:YMC-sols-1}
\end{figure}

\twocolumngrid

To address the characteristic time scales that are required for this mechanism to take place, let us consider a deviation from the exact partial solution, which describes the evolution of $U(t)$, and study numerically the general solution of the equations of motion --- see Appendix~\ref{Sec:EYM}. We first choose the subset of the initial conditions satisfying $Q_0\equiv Q(t=t_0)>1$, and then discuss the results of the numerical analysis qualitatively. For this choice of the initial conditions, Fig.~\ref{fig:YMC-sols-1} (left) illustrates the physical time evolution of the total energy density (in dimensionless units) of the homogeneous gluon condensate $U=U(t)$, namely $T^0_0 (t) \equiv \bar{\epsilon} + T^{0,{\rm U}}_0(t)$. In Appendix~\ref{Sec:EYM} we show the explicit expression of $T^{0,\rm U}_0$ and $\bar{\epsilon}$, respectively, as functionals of $U(t)$. In Fig.~\ref{fig:YMC-sols-1} (middle) we display the corresponding result for the trace of the total gluon EMT $T^\mu_\mu (t)\equiv 4\bar{\epsilon} + T^{\mu,\rm U}_\mu (t)$ in dimensionless units, and the corresponding solution for the logarithm of the scale factor is given in Fig.~\ref{fig:YMC-sols-1} (right).

\begin{figure}[!h]
\begin{minipage}{0.4\textwidth}
 \centerline{\includegraphics[width=1.0\textwidth]{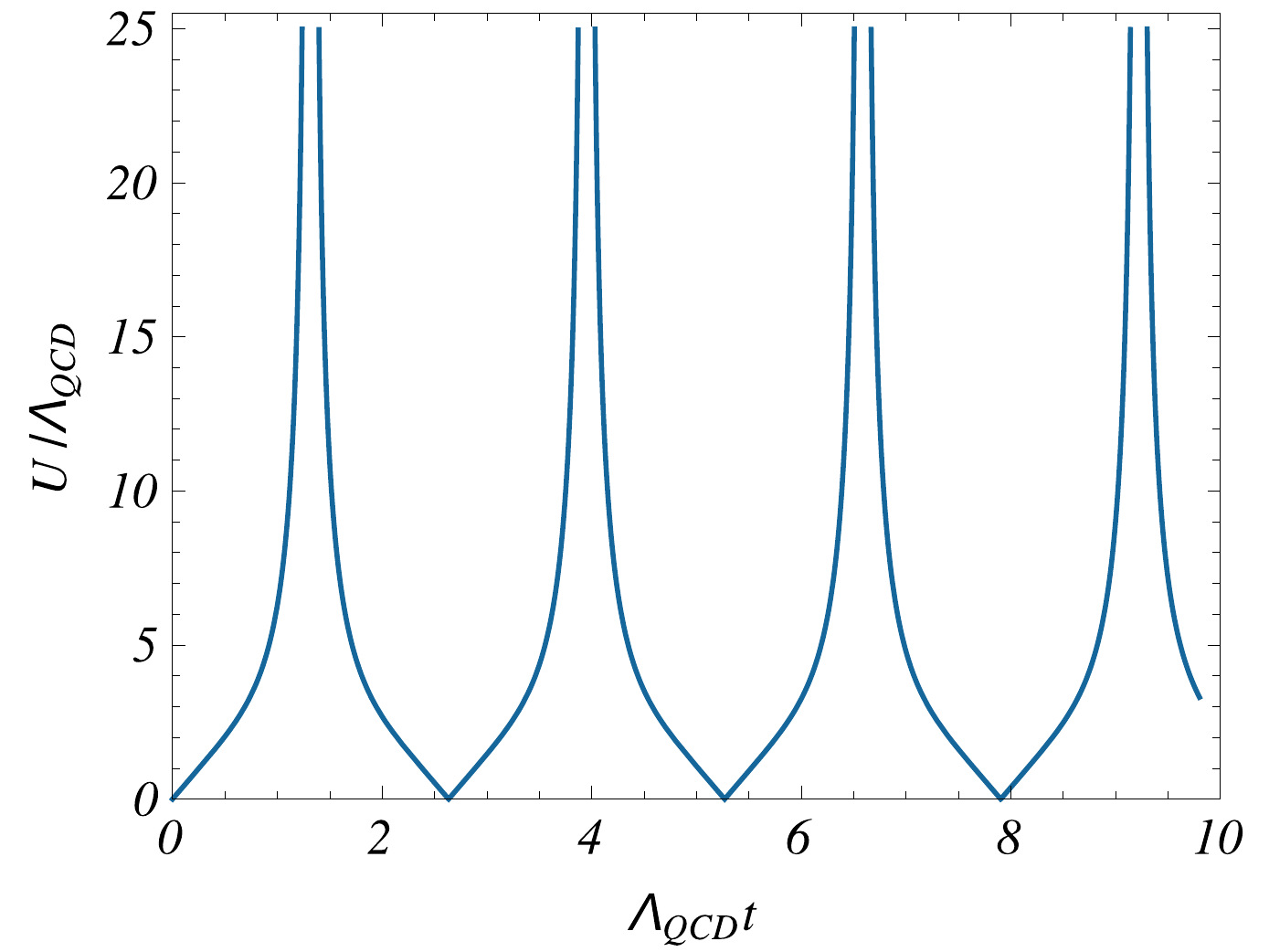}}
\end{minipage}
\hspace{1cm}
\begin{minipage}{0.4\textwidth}
 \centerline{\includegraphics[width=1.0\textwidth]{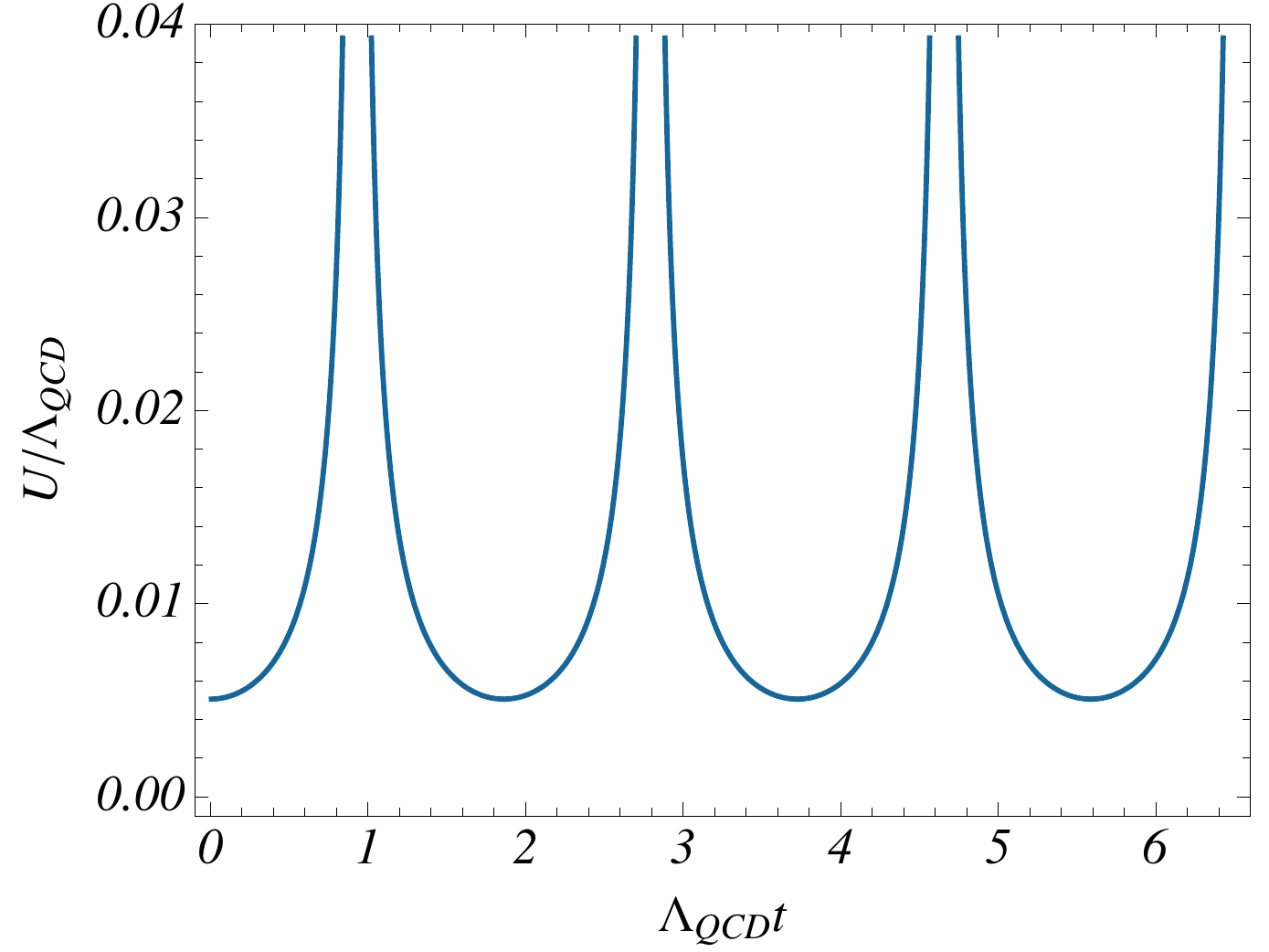}}
\end{minipage}
\caption{An illustration of the homogeneous QCD condensate amplitude oscillations $U=U(t)$, with quasi-periodic singularities in the physical time $t=\int ad\eta$, is shown for the chromoelectric ($Q(U)=1$) and the chromomagnetic ($Q(U)=-1$) vacua solutions, 
respectively in the left and right panels, in units of the characteristic time scale $\Lambda_{\rm QCD}^{-1}$. These spikes are localized in time-lapse, along the space-like directions, and must be interpreted as new solitonic 
solutions, dubbed {\it chronons} or {\it $\chi$-solutions}.}
\label{fig:YMC-sols}
\end{figure}

The period of the $T^\mu_\mu(t)$ oscillations is practically time independent, which can also be proven analytically, while a small residual time-dependence appears due to a possibly large deviation from $Q=1$. Here we used $\xi\simeq 4$ (following Ref.~\cite{Pasechnik:2013sga}) while a change of $\xi$ would only affect the asymptotic values of $T^0_0(t)$ and $T^\mu_\mu(t)$ at large $t$. Although the amplitude of the condensate $U(t)$ possesses quasi-periodic singularities, as is seen in Fig.~\ref{fig:YMC-sols}, the evolution of its energy density $T^{0,{\rm U}}_0(t)$ (see Fig.~\ref{fig:YMC-sols-1}, left), as well as of the pressure (or $T^{\mu,{\rm U}}_\mu(t)$, see Fig.~\ref{fig:YMC-sols-1}, middle), remain continuous in time. One immediately notices that the general solution asymptotically reaches a fix branch. This happens after a number of oscillations of the function $Q(t)$, whose amplitude approaches unity at large physical times $t$, i.e. $Q(t\to \infty)\Rightarrow 1$, for any initial conditions satisfying $Q_0>0$. During such a relaxation regime, the total energy density of the QCD vacuum (composed of the conventional QCD trace anomaly term -- the CM condensate -- and the considered CE homogeneous condensate) continuously decreases and eventually vanishes in the asymptotic limit $t\gg t_0$. Note, this regime is accompanied by a decelerating expansion of the Universe. 

The same quantities can be also studied for initial conditions of opposite sign, i.e. for $0<Q_0<1$. Numerical results for the latter regime are reported in Appendix~\ref{Sec:EYM}.  In this case the general solution asymptotically approaches the dS regime as well, in full analogy with the $Q_0>1$ case. A qualitatively similar situation is realized 
for $Q_0<0$. The dS (for $Q_0>0$) and AdS (for $Q_0<0$) solutions, therefore, appear as two attractor (or tracker) solutions of the Einstein-YM system, providing a dynamical mechanism for the elimination 
of the gluon vacuum component of the ground state energy of the Universe. This happens asymptotically, at macroscopic space-time scales and for arbitrary initial conditions 
and parameters of the model.

\onecolumngrid

\vspace{1cm}

\twocolumngrid

\section{Conclusions and remarks}
\noindent
We found an argument for the cancellation of the vacuum energy of QCD in the infrared limit that is related to the existence of an emergent $\mathbb{Z}_2$ Mirror symmetry of the RG flow --- derived from the bare Lagrangian. We showed that the cancellation of the vacuum energy is motivated both from the RG flow solutions and the effective action. We then commented on the relevance of vacuum fluctuations constituting chromoelectric vacuum solutions, which we argue counterbalance the conventional chromomagnetic contributions. We presented at this purpose an estimate of these latter contributions, and argued about the dynamical cancellation of the vacuum energy that they induce. We insisted that this is not in disagreement with lattice QCD. Our arguments are indeed dynamical, pertaining non-equilibrium configurations, and restricted to the infrared limit, to which we referred for the disappearance of the QCD vacuum beyond the confinement length-scale.

We claim that both the chromomagnetic and chromoelectric solutions co-exist in the infrared limit of QCD, while the topological effect is zero according to Ref.~\cite{Pagels}. Their attractor nature unavoidably leads to their compensation, and consequently induces a gross suppression of the net QCD vacuum contribution in the ground state of the Universe, naturally and without any fine tuning. While the quantum-topological effect may not be entirely excluded, the observed attractor nature of the QCD vacuum components may indicate similar self-tuning properties of the heterogeneous QCD vacuum, and point to a universal mechanism of the mutual net cancellation of all its components in the infrared limit of the theory, as required by cosmological observations. \\

Further comments on our analysis shall be pointed out:

\begin{itemize}  

\item
The appearance of negative values of $\bar{g}^2$ for the CE configuration in our analysis highlights a non-perturbative nature of the CE condensate and should not worry the reader about the appearance of ghosts. First of all, only gauge-invariant quantities were deployed in this analysis, including the operator $\mathcal{J}$, and thus both $\beta(\mathcal{J})$ and $\bar{g}^2(\mathcal{J})$. Nonetheless, the local loss of Lorentz invariance prevents from extending the ghost theorem to a case, like the one accounted for here, in which violations of the Lorentz symmetry --- in particular, of the boost invariance for the considered spatially homogeneous CE and CM configurations  --- appear locally. For instance, in the confined phase of YM theories the presence of CM vortices leads to a ground-state that is not spatially isotropic. This subtly allows to avoid the ghost theorem, which is formulated under the Lorentz invariance assumption. 
 
\item
It is commonly retained that the Savvidy's CE vacuum cannot be stable, since it is uniformly distributed in space (and has positive energy density). This statement is based on the old famous proof of Nielsen and Olsen (N.O.) in Ref.~\cite{NO}. {\it Mutatis Mutandis}, the same instability argument would imply the destabilization of the Mirror CM vacuum. However, the N.O. proof is not in contradiction with our results: we consider the evolution of the YM condensates in a {\it dynamical space-time}, whilst the N.O. argument is formulated on a rigid Minkowski space-time. This is enough to ensure that results reported in this analysis cannot be excluded on the basis of the N.O. proof. The N.O. instability is expected only after a relaxation time-scale of $\tau \simeq 50\div 100 \, \Lambda_{\rm QCD}^{-1}$ or similar --- see {\it e.g.}~Fig.~\ref{fig:YMC-sols-1}, where the dynamical evolution of the energy-momentum tensor of the condensate as a function of the cosmological time is displayed. The Savvidy's equations of motion for both the CE and CM condensates, at system with the Einstein's equations, provide a non-equilibrium fields system that was analyzed for the first time only within this work. Intriguingly, even starting from an initial non-zero energy-density, the evolution of the CE and CM YM condensates trigger a mutual screening, flowing towards a zero-energy density attractor. 

\item
In principle, after the cosmological relaxation time $\tau \simeq 50\div 100 \, \Lambda_{\rm QCD}^{-1}$, the Mirror Savvidy's vacuum states can decay into more complicated anisotropic vacuum states, in accordance with the N.O. argument. Nonetheless, the decay from a CE Savvidy's vacuum into a CM vacuum state --- either a CM Savvidy's vacuum or an anisotropic (averaged dominated) CM  vacuum state --- has to be exponentially suppressed. Indeed, since the CM and CE states are always separated by an energy barrier proportional to $\Lambda_{\rm QCD}$, at present times in the Universe, the tunnelling probability from the positive-energy CE vacuum to the CM minimal (negative) energy state is roughly $\Gamma \sim e^{-\Lambda_{\rm QCD}/T_{0}}\simeq 10^{-4 \times 10^{11}}\, $. This is not a surprising result, since a similar suppression of the tunnelling at present times also happens in the case of electroweak instantons (sphalerons). This argument allows the final anisotropic vacuum state to retain an energy density (averaged in  the $\sim \Lambda^{-3}$ confinement volume) that is vanishing. 

\item
The coexistence in the ``Fermi world'' of both the Mirror vacua can be sustained by non-perturbative and interpolating configurations. This is suggested by the well known case of a scalar field theory with a double wells potential: a kink-like profile can interpolate among two minima $-v$ and $v$ (with $v$ the VEV scale), which are contained in domain walls. It is very well known, for a scalar field related to the Higgs mechanism of a YM theory, that the change of the kink profile corresponds to the presence of monopole or vortex solutions that are localized inside the domain walls. Within the context of the YM theories, there is a strong evidence, sustained by numerical simulations, that the confinement phenomena are related by the formation of a network of 't Hooft monopoles or, alternatively, chromovortices --- see {\it e.g.}~Refs.~\cite{tHooft,Greensite:2003bk,Cornwall:1998ef,Ichie:1998na}. Both the scenarios highly suggest that these non-perturbative solutions may interconnect the CE and CM vacuum energies. In our case, the kink scalar profile may correspond to the effective $\mathcal{J}$-kink field. In other words, the cancellation mechanism proposed in our paper seems to naturally marry the confinement pictures that arise from numerical simulations. A fascinating picture to be envisaged consists into a sort of ``ferromagnetic confinement'' inside the Fermi world. In ferromagnetism, the ferro-material show several domain regions with certain different magnetic orientations. In a similar way, after the YM dimensional transmutation, several different domains, containing YM monopoles of vortices, interconnect the CE and CM vacua stabilised by the domain walls.

\item
In addition, the stabilization at asymptotically large times during the Universe's expansion of the ``false vacuum'', namely the positive-energy CE vacuum state, allows to reach an energy density that corresponds to the average energy of the two local minima. Decay of a ``false vacuum'' characterized by a positive cosmological constant, to a ``true vacuum'' can be achieved thanks to the Coleman de Luccia (CdL) instantonic solutions, and is exponentially suppressed. According to Ref.~\cite{Kanno:2011vm}, the CdL decay rate per unit time and unit volume can be exactly determined to be 
$$
\Gamma_{\rm CdL} \sim A e^{-48.33 \frac{\ell^2 }{\varkappa^2}} \!<\!\!<\! \ell^{-4}\,,
$$
where $A$ is a factor (irrelevant to this argument) derived from quantum corrections, $\ell$ is the typical length-scale of the system and $\varkappa^2 = 8\pi G$ is the Planck length squared. For the positive cosmological constant solution, the typical length-scale corresponds to the dS radius, {\it i.e.} $\ell \sim R_{\rm dS}$. It follows that the probability of decay from the ``false'' dS to the ``true'' AdS vacuum reads $$P_{\rm dec} \simeq R_{\rm dS}^4 \, \Gamma_{\rm CdL} \!<\!\!< \!1\,,$$
and that is generically very small --- see {\it e.g.} Ref.~\cite{Shlaer:2009vg}. These generic conditions actually make the ``false'' and positive-energy CE vacuum we dealt with in this analysis stable at all relevant cosmological time scales, also in full agreement with arguments given above.

\item
We may consider the case in which the Universe is filled with a positive-energy mean-field solution, which is such that the $\mathbb{Z}_2$ Mirror symmetry is broken for initial conditions (in general, away from the ground state) chosen at initial time $t=t_0$ in the early Universe. Consequently, the total YM field energy density is non-zero. This means that at $t=t_0$ any domain-wall has not formed yet for the chosen ($\mathbb{Z}_2$-violating) initial condition selected. Nonetheless, domain-walls will form later in time, due to the attractor nature of the solutions recovered here, and thanks to the $\mathbb{Z}_2$ symmetry that is restored globally at late times of the cosmological evolution. We then imagine the initial conditions such that the Universe expands, but that a generic mixed state is chosen that entails both positive-energy (CE) and negative-energy (CM) mean-fields, with a total energy density that is positive, but not yet corresponding to asymptotic tracker solutions and are away from the ground state (i.e. the ground-state of the effective YM action is not yet reached). This situation may correspond to a universe filled with radiation, for instance, the gluon plasma. At $t\!\sim\! t_0$, the CdL decay of the positive-energy CE configurations happens quite efficiently, such that the CE mean-field in a vicinity the ``false'' vacuum starts to decay, populating the CM ground state, which is the ``true'' vacuum. Due to local inhomogeneities, such decays do not happen at every space-time point with the same rate, but fluctuate from point to point, although carrying an averaged decay rate $\Gamma(t)$ that is generically a function of the cosmological time. For the case of QCD, the typical range of the gluon field fluctuations in the gluon plasma is around an inverse of the proton mass, more precisely $0.3$ fm, as known from particle phenomenology.

At some time $t^* \!>\!\!>\! t_0$, with $t^* \!<\!\!<\! \Gamma(t^*)^{-1}$, the Universe quickly expands, and both the initially large CE mean-field and the initially small CM mean-field, which appears due to the CdL decay, reach their attractors (i.e.~for each of the two configuration, the corresponding fields roll down to their respective minima of the effective Lagrangian), allowing for exactly opposite contributions to the cosmological constant. Nonetheless, this happens at different 3-space points regions, which on average are separated by length-scales of size $\sim0.3$ fm. This process then entails the emergence of spatially-separated stabilised patches of CM and CE vacua --- in other words, ``pockets'' of ``false'' and ``true'' vacua that correspond to dS and AdS attractor solutions necessarily emergent at $t^*\!>\!\!>t_0$. As we have elaborated in the point above, the patches of ``false'' vacuum with positive cosmological constant (dS solution) do stabilize and remain metastable with an extremely long life-time, way above the age of the Universe. We emphasize that the attractor solutions may not be reached simultaneously at every 3-space point, but that conversely at $t=t^*$ some small patches of space, with size $\sim 0.3 \, {\rm fm}$, may survive either in the CM condensate or in the CE condensate states, providing on average a vanishing contribution to the cosmological constant. In this sense, the $\mathbb{Z}_2$ symmetry is restored {\it on average} at $t^*$ when each of the patches in the YM system effectively reaches its cosmological attractor. At the same time, the presence of the $\mathbb{Z}_2$ symmetry between the different patches of solutions implies the formation of domain-walls between them as was mentioned above. The CdL decay basically terminates, and the patches get ``locked'' forever. This picture fits well with all we know (also experimentally) about the magnetic domains and the spontaneous magnetization in ferromagnetic materials strongly favouring the corresponding picture of confinement and provides its dynamical realisation in real physical time.

\item
Within the case of $\mathcal{N}\!=\!1$ SYM, described by the Veneziano-Yankielowicz (VY) superpotential, the dimensional transmutation is related to the gaugino condensation \cite{VY}. An initial $\mathbb{Z}_{N}$ symmetry, interpolating $N$ different vacua of the VY Lagrangian, is spontaneously broken down to $\mathbb{Z}_{2}$, due to the formation of domain walls. Since the VY and the Savvidy's Lagrangians are very similar, one might suggest that, in a non-SUSY model, the initial Mirror $\mathbb{Z}_{2}$ symmetry of the Savvidy's model can be violated by other non-perturbative configurations. Although this latter may represent an open possibility, nonetheless, according to simulations, the fact that the vacuum is dominated by the formation of YM monopole condensates or chromovortices seems to point exactly in the opposite direction. Indeed, these latter non-perturbative configurations preserve the $\mathbb{Z}_{2}$ Mirror symmetry of the vacua. Even more, such configurations may emerge due the Mirror symmetry. Even though their interactions may lead to a spontaneous symmetry breaking of the Mirror vacua, if we average on a confinement sphere with a finite radius $\Lambda_{\rm YM}^{-1}$, we expect that the averaged energy splitting among the Mirror vacua will still vanish. In this sense, the ground state of a non-SUSY YM theory may ``sit close'' to the correspondent SYM ones. 

\item
In Fig.~\ref{fig:YMC-sols} we have displayed an unexpected result, that the uniform parts of the CE and CM YM condensates form a periodic pattern of spikes in cosmological times. These latter show up as quantum ground-state solutions of the equations of motion for the CE and CM YM condensates, coupled to the Einstein's equation in the FLRW background, where the effective Lagrangian used as a classical field-theoretical model. The solutions are localized in time instants, but are not localized in 3-space, {\it i.e.} they cannot be interpreted as instantons. In other words, these solutions are space-like gauge solitons that appear and decay with a periodic time series. We were tempted to dub these solitons {\it chronons}, or {\it $\chi$-solutions}, as they likely remind S-branes within the context of string theory \cite{Gutperle:2002ai}. Chronons and anti-chronons trigger the dynamical screening mechanism depicted in Fig.~\ref{fig:YMC-sols-1}, dictating the dynamical flow to the net zero-energy density cosmological attractor. 

\item 
A further appealing possibility is that a non-perfect screening among the two CE and CM contributions may provide a source of dark energy, with the same dynamics analyzed in Refs.~\cite{Pasechnik:2013poa,Addazi:2016sot,Alexander:2016xbm,Alexander:2016xbm,Addazi:2016nok,Addazi:2016nok}, as an effect of a soft dynamical breaking of the Mirror symmetry (e.g.~in the first-order semiclassical gravity correction to the QCD ground state \cite{Pasechnik:2013poa}), while a screening mechanism of the Planckian contribution to the vacuum energy may arise from virtual black holes \cite{Addazi:2016jfq}.

\end{itemize}  

\acknowledgments
\noindent
Useful discussions with O. Teryaev, S. Brodsky, M. Faber, D. Antonov and D. Sivers are gratefully acknowledged. We would like to thank also G. Veneziano for a detailed 
constructive criticism on this subject. A.M.~wishes to acknowledge support by the Shanghai Municipality, through the grant No.~KBH1512299, and by Fudan University, through the grant 
No.~JJH1512105. R.P.~is supported in part by the Swedish Research Council grants, contract numbers 621-2013-4287 and 2016-05996, by CONICYT grants PIA ACT1406 and MEC80170112, 
as well as by the European Research Council (ERC) under the European Union's Horizon 2020 research and innovation programme (grant agreement No 668679). 
The work has been performed in the framework of COST Action CA15213 ``Theory of hot matter and relativistic heavy-ion collisions'' (THOR). This work was supported in part 
by the Ministry of Education, Youth and Sports of the Czech Republic, project LT17018.

\appendix

\section{YM ground state in the one-loop approximation}
\label{Sec:one-cond-1Lvsall}
\noindent
In this appendix we provide more computational details on the YM condensate dynamics on FLRW background. 

\subsection{Effective YM Lagrangian beyond one loop}
\label{Sec:two-cond}

\noindent
The all-loops effective action of SU(2) YM theory predicted by the FRG approach \cite{Dona:2015xia}, which can be {\it analytically} continued also to negative $\mathcal{J}<0$ as follows
\begin{eqnarray} \nonumber
\mathcal{L}_{\rm eff}&=&\frac{2\, \mathcal{J}}{16\pi^2}\int_0^{\infty}\frac{ds}{s}
\Big[e^{-s\sqrt{\frac{\lambda^4}{\mathcal{J} \tanh(\mathcal{J}/\lambda^{4}\varepsilon)}}}-e^{-s}\Big] \\
&\times& \Big[\frac{1}{4\sinh^2 s}-\frac{1}{4s^2}+1\Big] \,, \quad \varepsilon\ll 1 \,,
\label{FRG-L}
\end{eqnarray}
where 
\begin{eqnarray}
\label{an-cont}
\mathcal{J} \tanh\Big(\frac{\mathcal{J}}{\varepsilon\lambda^4}\Big)|_{\varepsilon\to0}\to |\mathcal{J}| \,.
\end{eqnarray}
This corresponds 
to the all-loops coupling constant
\begin{eqnarray} \nonumber
\big[\bar{g}^2\big]^{-1} &=& \frac{2}{4\pi^2}\int_0^{\infty}\frac{ds}{s}
\Big[e^{-s\sqrt{\frac{\lambda^4}{\mathcal{J} \tanh(\mathcal{J}/\varepsilon\lambda^4)}}}-e^{-s}\Big] \\
&\times& \Big[\frac{1}{4\sinh^2 s}-\frac{1}{4s^2}+1\Big] \,.
\label{FRG-g2}
\end{eqnarray}
By using the same trick as in Eq.~(\ref{an-cont}), one can analytically continue the one-loop results between 
the $\mathcal{J}>0$ and $\mathcal{J}<0$ domains as well.

Finally, the all-loops $\beta$-function can be found from the coupling above and it 
is conveniently written as
\begin{eqnarray} \nonumber
\frac{\beta}{\bar{g}^2}&=&\frac{-1}{2\pi^2}\sqrt{\frac{\lambda^4}{\mathcal{J} \tanh(\mathcal{J}/\varepsilon\lambda^4)}}
\int_0^{\infty}ds\,e^{-s\sqrt{\frac{\lambda^4}{\mathcal{J} \tanh(\mathcal{J}/\varepsilon\lambda^4)}}} \\
&\times& \Big[\frac{1}{4\sinh^2 s}-\frac{1}{4s^2}+1\Big] \,,
\label{FRG-beta}
\end{eqnarray}
which is an analytic function at any $\mathcal{J}$ for a finite $\varepsilon$. 

As was demonstrated in Ref.~\cite{Dona:2015xia}, by Taylor-expanding the second line of Eq.~(\ref{FRG-L}) 
for $s\to 0$ and keeping the first-order term as
\begin{eqnarray}
\Big[\frac{1}{4\sinh^2 s}-\frac{1}{4s^2}+1\Big] = \frac{11}{12} + {\cal O}(s^2) \,,
\end{eqnarray}
and then performing the integration with the help of the Schwinger formula
\begin{eqnarray}
\ln A = \int_0^{\infty}\frac{ds}{s} \Big[e^{-sA}-e^{-s}\Big] \,, \qquad A>0 \,,
\end{eqnarray}
one obtains the exact form of the one-loop effective action. 


\begin{figure}[!h]
 \centerline{\includegraphics[width=0.45\textwidth]{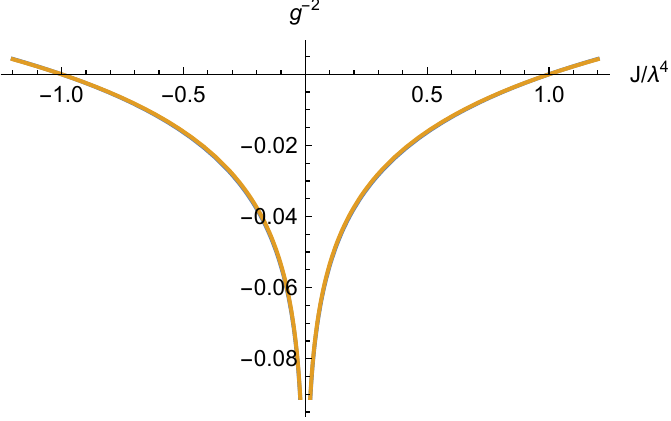}}
   \caption{ \small 
The running coupling $\bar{g}^{-2}(\mathcal{J})$ for the one-loop and all-loops cases 
in the effective $SU(2)$ theory ($\varepsilon=0.01$).}
 \label{fig:gm2}
\end{figure}

\begin{figure}[!h]
 \centerline{\includegraphics[width=0.45\textwidth]{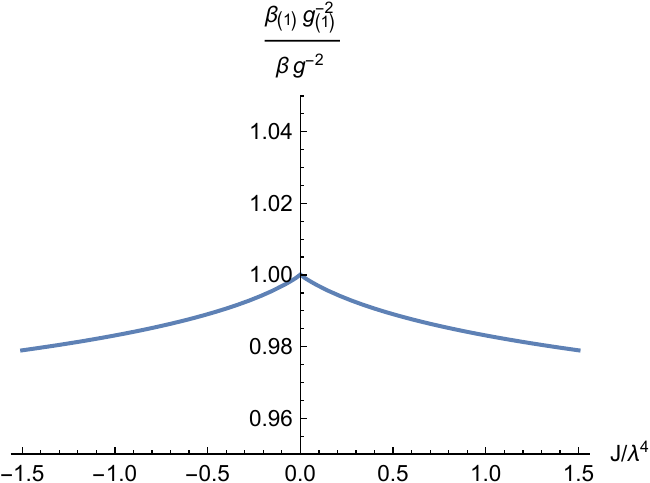}}
   \caption{ \small 
The ratio of the one-loop $\beta$-function coefficient $\beta_{(1)}/\bar{g}_{(1)}^2$ 
over the corresponding all-loops quantity given in Eq.~(\ref{FRG-beta}) ($\varepsilon=0.01$).
}
 \label{fig:beta-R}
\end{figure}


In the absence of gravity, the spatially homogeneous isotropic part  $\delta_{ik}U(t)$ of 
 satisfies the classical YM equations 
\begin{equation}
 (\dot{U})^2 + \bar{g}^2 \, U^4 = \mathrm{const} \,, \label{USU2}
\end{equation}
which can be integrated analytically  in terms of Jacobi Elliptic functions,
\begin{eqnarray} \nonumber
&& U(t) = U_0\, |{\rm cd}(\bar{g} U_0 t | -1)| \,, \\  
&& U(0)=U_0>0\,, \qquad \dot{U}(0)=0 \,.
\label{exactsolSU2}
\end{eqnarray}
The solution (\ref{exactsolSU2}) is an oscillatory quasi-harmonic\footnote{To a good approximation, 
the solution (\ref{exactsolSU2}) can be approximated by $U(t)\simeq U_0 {\rm cos}\left(\frac{2\pi}{T_U} t\right)$ 
\cite{Pasechnik:2013sga}.} function in $t$
\begin{eqnarray*}
&& U\big(n\, T_U\big) = U_0\,, \quad  
U\Big(\Big(n+\frac{1}{4}\Big)\, T_U\Big) = 0\,, \quad  n=0,1,2,\dots
\end{eqnarray*}
with amplitude $U_0>0$ and period
\begin{equation}
 T_U = B\left(\frac{1}{4},\frac{1}{2}\right)(\bar{g} U_0)^{-1}\simeq 1.2\times  \pi (\bar{g} U_0)^{-1} \,,
\end{equation}
where $B(x,y)$ is the Euler beta function. In the FLRW Universe, the classical YM condensate behaves as radiation medium 
such that $a(t)\propto t^{1/2}$, $p_{\rm YM}=\epsilon_{\rm vac}/3$ characteristic for the classical YM field behaviour.

\subsection{Condensate in the SU(3) YM theory}
\label{Sec:cond-SU3}

\noindent
The polar decomposition can be generalised to the $SU(3)$ YM theory 
\begin{equation}
 A_k=A_{a k} {\lambda_a\over 2}\,, \qquad \Big[{\lambda_a\over 2},\,{\lambda_b\over 2}\Big] =  
 if_{abc} {\lambda_c\over 2}\,, 
\end{equation}
using a special minimal $su(2)$ algebra embedding into the $su(3)$ in terms of three Gell-Mann matrices 
$\{\lambda_7,-\lambda_5,\lambda_2\}$, such that the resulting gauge-invariant symmetric field is linearly composed of components with well-defined 
transformation properties under spatial rotations. In particular, it has been shown, how to isolate 
the unique gauge-invariant and homogeneous/isotropic (spin-0) component.
Using this symmetric gauge approach, one straightforwardly generalises 
the gauge invariant $SU(2)$ decomposition of the gauge field into a unique spatially-homogeneous isotropic 
condensate and wave components to the $SU(3)$ case as follows 
\begin{eqnarray}
\!\!\!\!\!\!\!\!\!\! A_{ak} &=&
 \big( \delta_{a, 7} \delta_{k, 1} - \delta_{a, 5} \delta_{k, 2} + \delta_{a, 2} \delta_{k, 3}\big)U(t) + 
 \widetilde{A}_{ak} \,,
\label{SSU3}
\end{eqnarray}
for $a=1,\dots, 8$ and $\widetilde{A}_{ak}=\widetilde{A}_{ak}\big(t,\vec{x}\big)$. In the absence of gravity, the spatially homogeneous/isotropic gluon 
condensate $U(t)>0$ satisfies the classical YM equations
\begin{equation}
(\dot{U})^2 + \frac{1}{4}\, \bar{g}^2 \, U^4 = \mathrm{const}\,,
\label{USU3}
\end{equation}
Thus, we notice that the $SU(3)$ result (\ref{USU3}) for the condensate is the same as the $SU(2)$ result (\ref{USU2}), 
up to rescaling of the coupling constant as\footnote{We thank H.-P. Pavel for pointing out to us the ansatz
(\ref{SSU3}) and the simple rescaling of the coupling constant.} $\bar{g}\rightarrow \bar{g}/2$.

\section{One-condensate model: Einstein-YM equations of motion}
\label{Sec:EYM}

\noindent
By the variational principle, one obtains the EYM system of operator equations of motion in a non-trivial spacetime
\begin{eqnarray}
&&\frac{1}{\varkappa}\left(R_\mu^\nu-\frac12\delta_\mu^\nu R\right)=
\bar{\epsilon}\delta_\mu^\nu +
\frac{b}{32\pi^2}\frac{1}{\sqrt{-g}}\biggl[\biggl(-\mathcal{F}_{\mu\lambda}^a\mathcal{F}^{\nu\lambda}_a
\nonumber
\\
&& \quad +\,\frac14\delta_\mu^\nu
\mathcal{F}_{\sigma\lambda}^a\mathcal{F}^{\sigma\lambda}_a\biggr)
\ln\frac{e|\mathcal{F}_{\alpha\beta}^a\mathcal{F}^{\alpha\beta}_a|}{\sqrt{-g}\,
\lambda^4}-\frac14 \delta_\mu^\nu \,
\mathcal{F}_{\sigma\lambda}^a\mathcal{F}^{\sigma\lambda}_a\biggr]\,,\label{maineq} \\
&& \left(\frac{\delta^{ab}}{\sqrt{-g}}\overrightarrow{\partial}_\nu\sqrt{-g}-f^{abc}\mathcal{A}_\nu^c\right)
\left(\frac{\mathcal{F}_b^{\mu\nu}}{\sqrt{-g}}\,\ln\frac{e|\mathcal{F}_{\alpha\beta}^a
\mathcal{F}^{\alpha\beta}_a|}{\sqrt{-g}\,\lambda^4}\right)=0\,, \nonumber
\end{eqnarray}
where $\lambda\equiv \xi \Lambda_{\rm QCD}$ in terms of the QCD scale parameter $\Lambda_{\rm QCD}$ and an arbitrary scaling constant $\xi$, 
$e$ is the base of the natural logarithm, and the ground-state energy density
$\bar{\epsilon}\equiv \epsilon^{\rm QCD}_{\rm top}+\epsilon_{\rm CC}$
accounting for the quantum-topological QCD effect and the observable CC  only.

The system of equations of motion describing conformal time evolution of the gluon condensate $U=U(\eta)$ 
and the scale factor $a=a(\eta)$ reads
\begin{eqnarray}
&& \frac{6}{\varkappa} \frac{a''}{a^3} = 4\bar{\epsilon} + T^{\mu,{\rm U}}_{\mu}\,, \nonumber \\
&& T^{\mu,{\rm U}}_{\mu}=\frac{3b}{16\pi^2 a^4}\Big[(U')^2-\frac{1}{4}U^4\Big]\,, \nonumber \\
&& \frac{\partial}{\partial \eta}\Big(U'\,\ln\frac{6e\big|(U')^2-\frac{1}{4}U^4\big|}{a^4\lambda^4}\Big) \nonumber \\ 
&& \qquad +\,\frac{1}{2}U^3\,\ln\frac{6e\big|(U')^2-\frac{1}{4}U^4\big|}{a^4\lambda^4}=0 \,. \label{eqU}
\end{eqnarray}

The first integral of Eq.~(\ref{eqU}) is the Einstein $(0,0)$-equation and reads
\begin{eqnarray}
&& \frac{3}{\varkappa}\frac{(a')^2}{a^4}=\bar{\epsilon}+T^{0,{\rm U}}_{0}\,, \nonumber \\
&& T^{0,{\rm U}}_{0}=\frac{3b}{64\pi^2 a^4}\,\Big(\Big[(U')^2+\frac{1}{4}U^4\Big]\,
\ln\frac{6e\big|(U')^2-\frac{1}{4}U^4\big|}{a^4\lambda^4} \nonumber \\
&& \qquad +\, (U')^2 - \frac{1}{4}U^4\Big)\,. \label{eqUint}
\end{eqnarray}

\subsubsection{The general asymptotic solutions of the EYM system}
\label{Sec:Gsol}

Omitting ordinary matter components in the cosmological plasma, the system of EYM equations (\ref{eqU}), (\ref{eqUint}) 
in physical time reads
\begin{eqnarray}
\frac{6}{\varkappa}\Big[\frac{\ddot{a}}{a} + \frac{\dot{a}^2}{a^2}\Big] = 4\bar{\epsilon} + T^{\mu,{\rm U}}_{\mu}\,, \quad 
\frac{3}{\varkappa}\frac{\dot{a}^2}{a^2}=\bar{\epsilon} + T^{0,{\rm U}}_{0}\,, 
\label{eqfirst}
\end{eqnarray}
where the energy desity of the gluon condensate and the trace in the one-loop effective YM theory read
\begin{eqnarray}
&& T^{0,{\rm U}}_{0} = \frac{33}{64\pi^2 a^4}\Big(\Big[a^2\dot{U}^2+\frac{1}{4}U^4\Big]
\ln \frac{6e\big|a^2\dot{U}^2-\frac{1}{4}U^4\big|}{a^4(\xi \Lambda_{\rm QCD})^4} \nonumber \\ 
&& \qquad +\, a^2\dot{U}^2 - \frac{1}{4}U^4\Big)\,, \nonumber \\
&& T^{\mu,{\rm U}}_{\mu} = \frac{33}{16\pi^2 a^4}\Big[a^2\dot{U}^2-\frac{1}{4}U^4\Big]\,,
\label{PTstuff}
\end{eqnarray}
respectively. Since $U(t)$ has quasi-periodic singularities, it is more convenient to work in terms of a new continuous universal 
function $g=g(t)$ defined as
\begin{eqnarray}
&& T^{\mu,{\rm U}}_{\mu} - C = (g(t)+1)\Big[T^{0,{\rm U}}_0 - \frac{C}{4} \Big]\,, \nonumber \\
&& C \equiv - 4\epsilon^{\rm QCD}_{\rm top}=\frac{33}{16 \pi^2}\frac{(\xi \Lambda_{\rm QCD})^4}{6e}\,,
\label{defg}
\end{eqnarray}
such that the equations (\ref{eqfirst}) can be written explicitly in terms of continuous functions
\begin{eqnarray}
&& \frac{6}{\varkappa}\Big[\frac{\ddot{a}}{a}+\frac{\dot{a}^2}{a^2}\Big]=4\epsilon_{\rm CC} + 
(g(t)+1) \Big[T^{0,{\rm U}}_{0} - \frac{C}{4}\Big]\,, \nonumber \\
&& \frac{3}{\varkappa}\frac{\dot{a}^2}{a^2}=\epsilon_{\rm CC}- \frac{C}{4} + T^{0,{\rm U}}_0\,, \;\; 
T^{0,{\rm U}}_0=T^{0,{\rm U}}_0(U,\dot{U},a) \,,
\label{gT}
\end{eqnarray}
where the last equation is given by Eq.~(\ref{eqfirst}). Excluding $T^{0,{\rm U}}_{0}$ we arrive at the 
equation for the scale factor
\begin{eqnarray}
6\frac{\ddot{a}}{a}+3(1-g(t))\frac{\dot{a}^2}{a^2}+\varkappa\epsilon_{\rm CC}(g(t)-3)=0\,,
\label{eqa}
\end{eqnarray}
whose general solution has the following form

\onecolumngrid

\begin{eqnarray}
&& a(t) = a^* \, \exp\Big[\sqrt{\frac{\varkappa \epsilon_{\rm CC}}{3}}\times \nonumber \\
&& \int_{t_0}^t\frac{1+\sqrt{\frac{\epsilon_{\rm CC}}{\epsilon_0}}+
\Big(1-\sqrt{\frac{\epsilon_{\rm CC}}{\epsilon_0}}\Big)\,\exp\Big\{\sqrt{\frac{\varkappa \epsilon_{\rm CC}}{3}}
\Big(-3(t'-t_0)+\int_{t_0}^{t'}g(\tau)d\tau\Big)\Big\}}{1+\sqrt{\frac{\epsilon_{\rm CC}}{\epsilon_0}}-
\Big(1-\sqrt{\frac{\epsilon_{\rm CC}}{\epsilon_0}}\Big)\,
\exp\Big\{\sqrt{\frac{\varkappa \epsilon_{\rm CC}}{3}}\Big(-3(t'-t_0)+\int_{t_0}^{t'}g(\tau)d\tau\Big)\Big\}}dt' \Big] \,,
\label{aexact}
\end{eqnarray}

\twocolumngrid

in terms of the initial values of the scale factor $a^*\equiv a(t=t_0)$ and the total energy density $\epsilon_0$ , 
respectively. Besides, the total energy density and the trace of the EMT found in Eqs.~(\ref{eqUint}) 
and (\ref{eqU}), respectively, as functions of physical time read

\onecolumngrid

\begin{eqnarray}
&& \frac{T^0_0(t)}{\epsilon_{\rm CC}}=\left[\frac{1+\sqrt{\frac{\epsilon_{\rm CC}}{\epsilon_0}}+
\Big(1-\sqrt{\frac{\epsilon_{\rm CC}}{\epsilon_0}}\Big)\,\exp\Big\{\sqrt{\frac{\varkappa \epsilon_{\rm CC}}{3}}
\Big(-3(t-t_0)+\int_{t_0}^{t}g(\tau)d\tau \Big)\Big\}}{1+\sqrt{\frac{\epsilon_{\rm CC}}{\epsilon_0}} - 
\Big(1-\sqrt{\frac{\epsilon_{\rm CC}}{\epsilon_0}}\Big)\,\exp\Big\{\sqrt{\frac{\varkappa 
\epsilon_{\rm CC}}{3}}\Big(-3(t-t_0)+\int_{t_0}^{t}g(\tau)d\tau\Big)\Big\}}\right]^2  \label{ETexact} \\
&& \frac{T^{\mu}_{\mu}(t)}{\epsilon_{\rm CC}} = 4 + \frac{4(g(t)+1)\Big(1-\frac{\epsilon_{\rm CC}}
{\epsilon_0}\Big)\,\exp\Big\{\sqrt{\frac{\varkappa \epsilon_{\rm CC}}{3}}
\Big(-3(t-t_0)+\int_{t_0}^{t}g(\tau)d\tau\Big)\Big\}}{\left[1+\sqrt{\frac{\epsilon_{\rm CC}}{\epsilon_0}}-
\Big(1-\sqrt{\frac{\epsilon_{\rm CC}}{\epsilon_0}}\Big)\,
\exp\Big\{\sqrt{\frac{\varkappa \epsilon_{\rm CC}}{3}}\Big(-3(t-t_0)+\int_{t_0}^{t}g(\tau)d\tau\Big)\Big\}\right]^2}\,.
\nonumber 
\end{eqnarray}

\twocolumngrid

Here $\epsilon_0$ is initial energy density of the gluon condensate, and $\epsilon_{\rm CC}\ll \epsilon_0$ is 
the observed cosmological constant. 

Note, the relations (\ref{aexact}) and (\ref{ETexact}) do not rely on any approximations and provide 
the general solution of the equations (\ref{eqfirst}) as long as a solution for $g(t)$ is known.
We search the latter solution in a close vicinity of the exact one where 
$Q_0\equiv Q(t=t_0)\sim 1$. Then expanding the YM energy density around asymptotic value 
given by the exact solution, $T^{0,{\rm U}*}_{0}\equiv C/4$, we write
\begin{eqnarray}
T^{0,{\rm U}}_{0}(t) \simeq C/4 + \delta\epsilon(t)\,, \qquad \delta\epsilon \ll C \,.
\label{decompH}
\end{eqnarray}
The time derivatives $\dot{a}$ and $\dot{T}^{0,{\rm U}}_0$ take two different asymptotic forms depending
on relation between $\delta\epsilon(t)$ and $\epsilon_{\rm CC}$. In particular, in the case of large 
$\delta\epsilon(t)\gg \epsilon_{\rm CC}$ we get
\begin{eqnarray}
&& \dot{a} \simeq \sqrt{\frac{\varkappa}{3}}\,a\,\sqrt{\delta\epsilon}\,, \nonumber \\
&& \dot{T}^{0,{\rm U}}_{0} \simeq \sqrt{\frac{\varkappa}{3}}(g-3)(\delta\epsilon)^{3/2}\,,
\label{decompda1}
\end{eqnarray}
keeping only the leading order terms in $\delta\epsilon \ll C$. In the opposite case when 
$\delta\epsilon(t)\ll \epsilon_{\rm CC}$, we obtain
\begin{eqnarray}
&& \dot{a} \simeq \sqrt{\frac{\varkappa\epsilon_{\rm CC}}{3}}\,a\,\Big(1+\sum_{n=1}^{\infty}{\frac{1}{2} \choose n}
\Big(\frac{\delta\epsilon}{\epsilon_{\rm CC}}\Big)^n\Big)\,,\nonumber \\
&&\!\!\!\!\!\!\!\!\!\!\!\!\!\!\dot{T}^{0,{\rm U}}_{0} \simeq \sqrt{\frac{\varkappa\epsilon_{\rm CC}}{3}}(g-3)\,\delta\epsilon\, 
\Big(1+\sum_{n=1}^{\infty}{\frac{1}{2} \choose n}\Big(\frac{\delta\epsilon}{\epsilon_{\rm CC}}\Big)^n\Big)\,.
\label{decompda2}
\end{eqnarray} 
It will be shown later that the difference between Eqs.~(\ref{decompda1}) and (\ref{decompda2}) is not relevant 
and leads to a very small correction to the period of oscillations of $g=g(t)$ which can be neglected for practical purposes.
\begin{figure}[!h]
 \centerline{\includegraphics[width=0.45\textwidth]{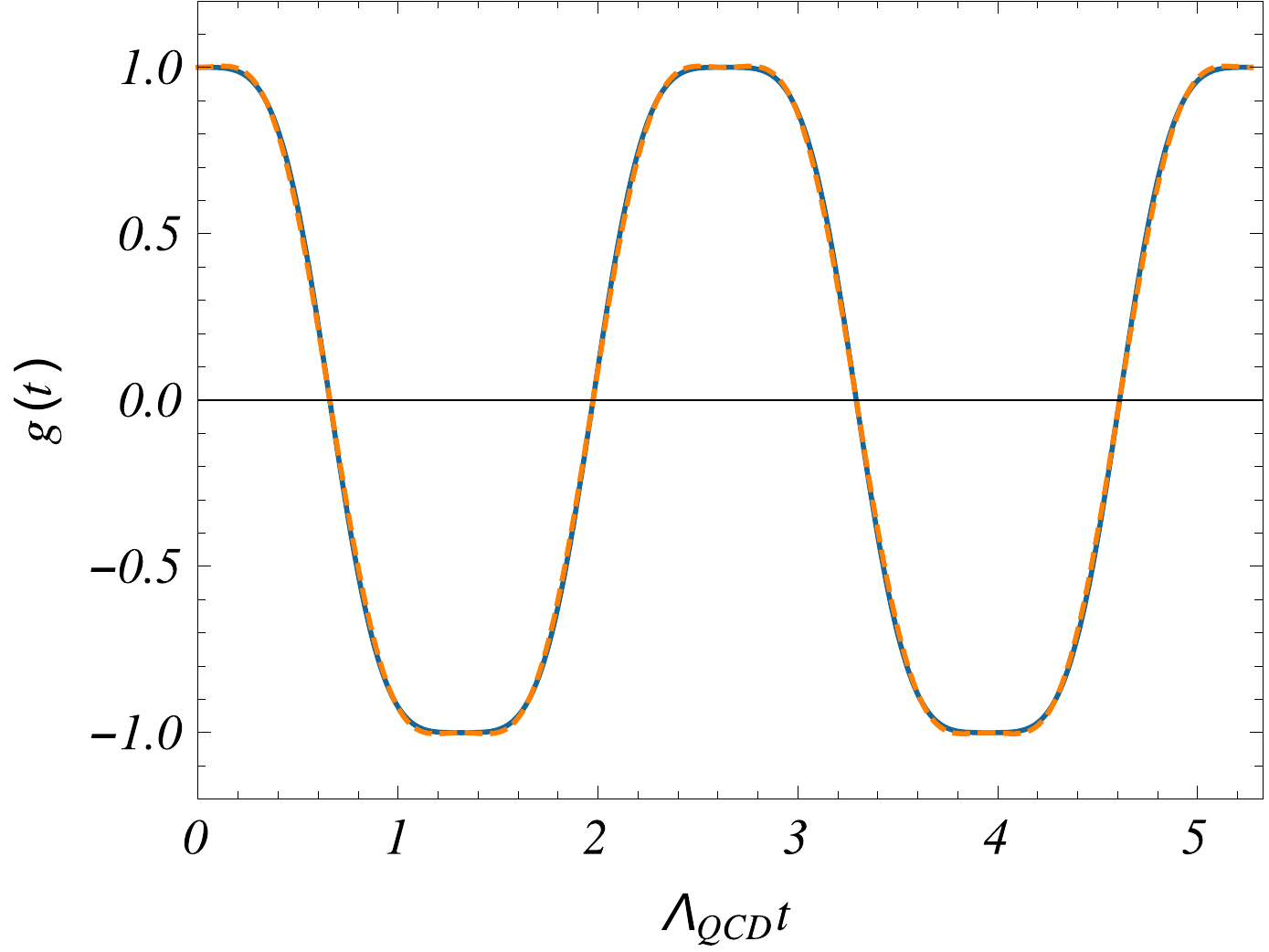}}
   \caption{The time dependence of the quasi-harmonic universal function $g=g(t)$ corresponding to 
   the exact solution (\ref{gsol}) (solid line) compared to its Fourier approximation (\ref{approx}) (dashed line).}
\label{fig:gtplot}
\end{figure} 

Using Eqs.~(\ref{eqfirst}) and (\ref{defg}) with expansions (\ref{decompH}) and (\ref{decompda1}) in the case of large 
$\delta\epsilon(t)\gg \epsilon_{\rm CC}$, we finally arrive at the equation for $g=g(t)$
\begin{eqnarray}
\dot{g}^4 - \frac{8(\xi \Lambda_{\rm QCD})^{4}}{3e}\,(1-g^2)^3 = 0 \,.
\label{eqgmain}
\end{eqnarray} 
Its implicit analytic solution can be found for the inverse function $0<t(g)<T_g/2$ within 
the period of oscillations $T_g$ of $g(t)$ function as
\begin{eqnarray}
 && t(g) = -\frac{(6e)^{1/4}}{2\xi\Lambda_{\rm QCD}}\,
 \Big[{}_2F_1\Big(\frac{1}{2},\frac{3}{4};\frac{3}{2};g^2\Big)g-k\Big]\,, \label{Teta} \\ 
 && \qquad \quad 0<t(g)<T_g/2 \,,
\label{gsol}
\end{eqnarray}
where the constant $k=2.622$ is defined by Eq.~(\ref{Teta}), $T_g$ is the period of oscillations of the $g(t)$ function, 
and the initial condition $g(t=0)=1$ is adopted, for simplicity. Under these conditions Eq.~(\ref{gsol}) determines $g=g(t)$ 
as a periodic quasi-harmonic function with unit amplitude. Then, its period is straightforwardly found as
\begin{eqnarray}
T_g=\frac{2(6e)^{1/4}}{\xi\Lambda_{\rm QCD}}\int_{0}^1\frac{dg}{(1-g^2)^{3/4}}=\frac{2 k (6e)^{1/4}}{\xi \Lambda_{\rm QCD}}\,,
\label{period}
\end{eqnarray} 
which is two times less the period of the YM condensate oscillations $U(t)$ and coincides with naive expectation (\ref{gT}).

Now, repeating the above calculation in the opposite case corresponding to $\delta\epsilon(t)\ll \epsilon_{\rm CC}$
with Eq.~(\ref{decompda2}) to the equation similar to Eq.~(\ref{eqgmain}) but with a small additional term $\propto \alpha\ll 1$
\begin{eqnarray}
&& \frac{dg}{d\tau} \pm 2(1-g^2)^{3/4}-\alpha(1-g^2)=0\, ,\qquad \tau=t\,\frac{\xi\Lambda_{\rm QCD}}{(6e)^{1/4}}\,, \nonumber \\
&& \qquad T_g' \simeq T_g\,\Big[1+\frac{\pi}{4k^2}\alpha^2\Big]\,,\nonumber \\ 
&& \qquad \alpha =\frac{2(6e)^{1/4}}{\xi\Lambda_{\rm QCD}}\, 
\sqrt{\frac{\varkappa\epsilon_{\rm CC}}{3}}\approx 4.7\times 10^{-42} \,.
\label{secondwayeq}
\end{eqnarray}

Note, the function $g(t)$ satisfies the following integral constraint
\begin{eqnarray}
&& \int_0^t g(\tau)d\tau = \pm\frac{(6e)^{1/4}}{\xi\Lambda_{\rm QCD}}(1-g^2)^{1/4}\,, \nonumber \\
&& \qquad \frac{nT_g}{2} < t < \frac{(n+1)T_g}{2} \,,
\label{int}
\end{eqnarray}
where the upper sign corresponds to even $n$ and the lower -- to odd $n$. The constraint (\ref{int})
can further be used in Eqs.~(\ref{aexact}) and (\ref{ETexact}) in order to express the general 
EYM solutions for $a(t)$, $T^0_0(t)$ and $T^{\mu}_{\mu}(t)$ in terms of $g(t)$.
A good analytic approximation to the exact $g(t)$ solution (\ref{gsol}) can be constructed 
keeping only the first two non-vanishing harmonic Fourier-terms, namely,
\begin{eqnarray}\label{approx}
&& g(t) \simeq A \cos\Big(\frac{2\pi t}{T_g}\Big) + (1-A) \cos\Big(\frac{6\pi t}{T_g}\Big)\,,\\
&& A=\frac{2}{k}\int_0^1\frac{g}{(1-g^2)^{3/4}}\,\cos\Big(\frac{\pi}{2k}\int_g^1\frac{dx}{(1-x^2)^{3/4}}\Big)dg \approx 1.14\,.
\nonumber 
\end{eqnarray}
In Fig.~\ref{fig:gtplot} we observe that the formula (\ref{approx}) approximates the exact 
solution for the universal $g(t)$ function found from Eq.~(\ref{gsol}) with a very good accuracy.


\end{document}